\documentclass[12pt]{article}
\usepackage{amsfonts,amsmath,amssymb}
\usepackage{bm}
\usepackage{bbm}
\usepackage[pdftex]{graphicx}
\usepackage{hyperref}
\usepackage{float}
\usepackage{ltxtable}
\newcommand{\pa}{\partial}
\newcommand{\nn}{\nonumber}
\newcommand{\Ord}{{\cal O}}

\makeatletter

\@addtoreset{equation}{section}
\makeatother

\def\href#1#2{#2}
\begin{document}

\begin{titlepage}

\begin{center}

\hfill 
\vskip32mm

\textbf{\Large Extra-Natural Inflation (De)constructed}\\[16mm] 

{\large Kazuyuki Furuuchi, Noel Jonathan Jobu}\\[2mm]
{\large and}\\[2mm] 
{\large Suvedha Suresh Naik}\\[10mm]

{\sl
{Manipal Centre for Natural Sciences}\\
{Centre of Excellence}\\
{Manipal Academy of Higher Education}\\
{Dr.~T.M.A.~Pai Planetarium Building}\\
{Manipal 576 104, Karnataka, India}
}

\vskip8mm 
\end{center}
\begin{abstract}
Extra-natural inflation is (de)constructed.
Explicit models
are compared with cosmological observations.
The models successfully achieve trans-Planckian
inflaton field excursions.
\end{abstract}

\end{titlepage}

\tableofcontents

\section{Introduction}\label{sec:intro}

Dimensional (de)construction
\cite{ArkaniHamed:2001ca,Hill:2000mu}
provides purely $4$D QFT description of latticized extra dimensions.
(De)construction of the gauge-Higgs unification model 
\cite{Hosotani:1983xw,Hosotani:1988bm,Davies:1988wt,%
Antoniadis:1990ew,Hatanaka:1998yp}
has provided a new mechanism to protect the Higgs mass
against quantum corrections \cite{ArkaniHamed:2001nc}.
Many mechanisms which were used to explain
the lightness of the Higgs mass
have also been used to
explain the flatness of the potential of
slow-roll inflation models.
In the case of the gauge-Higgs unification model,
extra-natural inflation 
\cite{ArkaniHamed:2003wu,Kaplan:2003aj}
employs the same mechanism in slow-roll inflation
and provides a microscopic theory
of natural inflation \cite{Freese:1990rb}
from extra dimensions.
Given the (de)construction 
of the gauge-Higgs unification model,
it is natural to explore (de)construction of
extra-natural inflation.
However, already in the original work 
\cite{ArkaniHamed:2003wu},
it has been noticed that
(de)construction of extra-natural inflation
with one (de)constructed extra dimension
does not lead to a successful model of slow-roll inflation.
The obstacle was as follows:
The basic natural inflation model
is a large field inflation model
which is required to have
a trans-Planckian inflaton field excursion
to explain observations of
Cosmic Microwave Background (CMB) anisotropy.
The trans-Planckian inflaton excursion
requires $2 \pi {F} \gg M_P$,
where ${2\pi F}$ is the period of the inflaton potential and
$M_P$ is the reduced $4$D Planck scale.
However, in (de)construction models
with one (de)constructed extra dimension,
${F}$ is related to a symmetry breaking scale
$f$ in the model
as ${F} = f / \sqrt{N}$,
where 
$N$ is the number of the lattice points
in the (de)constructed dimension.
For the model to be 
described
without taking into account 
strong quantum gravity effects,
$f \ll M_P$ is required,
leading $F \ll M_P$.
This poses an obstacle for 
having the
trans-Planckian inflation excursion.
Thus nearly two decades after
(de)construction was proposed,
there has been no notable application
of it in inflation model building.
However, recently
two ways to circumvent the above obstruction
have been found \cite{Furuuchi:2020klq}.
One is to introduce 
a gauge-invariant Stueckelberg potential
which gives rise to the dominant part of the inflaton potential.
The gauge invariant Stueckelberg potential
is not periodic
under the shift of the inflaton,
which originates from 
the extra-dimensional component of the gauge field.%
\footnote{Since the periodicity of the action
originates from the gauge invariance,
it must be an exact symmetry.
However, the gauge symmetry transformation
involves the transformation of the Stueckelberg field.
Thus the potential is not periodic if only
the gauge field is shifted \cite{Furuuchi:2015foh}.}
Therefore, the inflaton field excursion is not restricted by 
the periodicity ${2\pi F}$.
Another way 
to circumvent the obstruction
is to increase the number of 
(de)constructed extra dimensions.
It was shown 
in \cite{Furuuchi:2020klq}
that the period ${2\pi F}$ of the inflaton
is related to the symmetry breaking scale $f$
as ${F} = f N^{\frac{d}{2}-1}$,
where $d$ is the number of the (de)constructed extra dimensions
and $N$ is the number of the lattice points 
in each direction.%
\footnote{Here, for simplicity, 
the number of the lattice points in all 
(de)constructed dimensions are chosen to be the same.
In the main body, we will use
an improved choice of parameters 
which makes the $d=2$ case also worth examining.}
Therefore, for $d \geq 3$, the period ${2\pi F}$ of the 
inflaton potential can be much larger 
than the symmetry breaking scale $f$ 
if $N$ is sufficiently large,
and this may enable the
trans-Planckian inflaton excursion.
In \cite{Furuuchi:2020klq}, the first way was the main focus,
while the second way was briefly mentioned.
In this article, we will study the second way 
in more detail.
We construct explicit inflation models
in which the zero-mode of 
a gauge field 
in one of the (de)constructed direction
is an inflaton,
and the field range of the inflaton 
is enhanced by (de)construction
to enable the trans-Planckian excursion.
Then we study the constraints 
on the parameters of the models
from CMB observations.

The organization of this article is as follows:
In Sec.~\ref{sec:model}
we present
the theoretical framework of the
(de)construction of extra-natural inflation.
We start with a high energy theory
with a product gauge group,
and derive the low energy effective action
which is
appropriate below the energy scale of
the gauge symmetry breaking
to the diagonal subgroup.
One of the (de)constructed extra dimensional components
of the gauge field is to be identified with the inflaton.
The one-loop effective potential for the inflaton is derived.
The charged matter field contents 
determine the one-loop effective potential.
In Sec.~\ref{sec:obs}
we compare the models with explicit choice
of the charged matter field contents
with CMB observations.
With charged matter fields having different charges,
the (de)construction models
provide microscopic theories
of a version of natural inflation called
multi-natural inflation \cite{Czerny:2014wza},
which can explain
the latest CMB observations well.
The observational 
constraints on the model parameters are derived.
In a region of the model parameters,
our (de)constructed models of 
extra-natural inflation successfully 
describe large-field inflation.
We conclude with summary and discussions
in Sec.~\ref{sec:summary}.
Some useful formulas and technical details
are collected in the Appendices.

\section{Extra-natural inflation 
(de)constructed}\label{sec:model}

The theoretical framework for 
(de)constructing
extra-natural inflation have already been
developed
in \cite{Furuuchi:2020klq},
which studied (de)construction of a massive gauge theory.
More detailed calculations and explanations 
are given there,
and interested readers are 
encouraged to read the above reference together.

The (de)constructed extra dimensions 
we consider will be
a $d$-dimensional periodic lattice
(a lattice on a $d$-dimensional torus)
with $N_{I}$ 
(${I} = 1,2,\cdots,d$)
lattice points
in the ${I}$-th direction.
The model with such (de)constructed extra dimensions
is described by the following $4$D action:
\begin{align}
S_{(4+d)}
=
\int d^4 x
\sum_{\vec{j}}
\Biggl\{
&- \frac{1}{4} F_{\mu\nu (\vec{j})}  F^{\mu\nu}_{(\vec{j})} 
\Biggr.
+ 
\sum_{I=1}^d
\frac{f_I^2}{2}
D_\mu U_{(\vec{j},\vec{j}+\vec{e}_I)}^I D^{\mu} U_{(\vec{j},\vec{j}+\vec{e}_I)}^{I\dagger}
\nn\\
&
+
\frac{1}{2}
\beta_{IJ}
g^2 f_I^2 f_J^2\,
\mathrm{Re}
\left[
U_{(\vec{j},\vec{j}+\vec{e}_I)}^I 
U_{(\vec{j}+\vec{e}_I,\vec{j}+\vec{e}_I+\vec{e}_J)}^J
U_{(\vec{j}+\vec{e}_J,\vec{j}+\vec{e}_I+\vec{e}_J)}^{I\dagger}
U_{(\vec{j},\vec{j}+\vec{e}_J)}^{J\dagger}
\right]
\nn\\
&
+
{\cal L}_{matter}
+
\ldots
\Biggl.\Biggr\}\,,
\nn\\
(\mu &= 0,1,2,3;\, 
I = 1,2,\cdots,d;\, 
j_I=0,1,\cdots,N_I-1 \,\, \mathrm{mod} \,\, N_I).
\label{eq:decS4pd}
\end{align}
Here, ``$\ldots$'' 
represent higher dimensional operators
which are irrelevant at low energy.
The matter Lagrangian density
${\cal L}_{matter}$ 
will be specified later.
The $I$-th component of $\vec{j}$
is denoted as $j_I$.
The $d$-dimensional vector $\vec{j}$ 
parametrizes the lattice points.
$\vec{e}_{I}$ is a vector 
whose ${J}$-th component is given by $\delta_{{I} {J}}$.
The field $U_{(\vec{j},\vec{j}+\vec{e}_{I})}^I$
can be regarded as a parametrization of the
Nambu-Goldston boson from a global $U(1)$ symmetry breaking
with the symmetry breaking scale $f_I$
\cite{ArkaniHamed:2001ca}.
At the same time, in the language of lattice gauge theory,
it is a link variable connecting
the lattice points $\vec{j}$ and $\vec{j}+\vec{e}_{I}$.
This lattice (de)constructs effective extra dimensions
from purely 4D QFT.
The link variables can be parametrized as
\begin{equation}
U_{(\vec{j},\vec{j}+\vec{e}_{I})}^{{I}}
=
\exp
\left[
i \frac{A_{(\vec{j},\vec{j}+\vec{e}_{I})}^{I}}{f_{I}}
\right]\,.
\label{eq:Ualpha}
\end{equation}
In the language of $4$D QFT,
the field $A^I_{(\vec{j},\vec{j}+\vec{e}_{I})}$ 
is analogous to the pion field,
which is the approximate Nambu-Goldstone boson
from the chiral symmetry breaking.
At the same time, in the language of lattice gauge theory,
the fields $A^I_{(\vec{j},\vec{j}+\vec{e}_{I})}$'s make up
the gauge field in the (de)constructed directions.

The action \eqref{eq:decS4pd} 
has the product $\prod_{\vec{j}} U(1)_{(\vec{j})}$ gauge symmetry.
We also impose the symmetry under 
the discrete translation:
\begin{equation}
\vec{j} \rightarrow \vec{j} + \vec{e}_I
\qquad (I = 1,2,\cdots,d)\,,
\label{eq:DT}
\end{equation}
so that the gauge coupling $g$ is the same
for all $U(1)_{(\vec{j})}$.
The gauge transformation generated by
$g_{(\vec{j})}(x) = e^{i g \alpha_{(\vec{j})}(x)}$ are given as
\begin{align}
A_{\mu (\vec{j})} (x) 
&\rightarrow 
A_{\mu (\vec{j})}(x) - \pa_\mu \alpha_{(\vec{j})}(x)\,,
\\
U_{(\vec{j},\vec{j}+\vec{e}_I)}^I (x) 
&\rightarrow 
g_{(\vec{j})}^{-1}(x) U_{(\vec{j},\vec{j}
+\vec{e}_I)}(x) g_{(\vec{j}+\vec{e}_I)}(x)\,.
\end{align}
The covariant derivative
in \eqref{eq:decS4pd}
is defined as
\begin{equation}
D_\mu U_{(\vec{j},\vec{j}+\vec{e}_I)}^I
=
\pa_\mu U_{(\vec{j},\vec{j}+\vec{e}_I)}^I
- i g A_{\mu (\vec{j})} U_{(\vec{j},\vec{j}+\vec{e}_I)}^I
+ i g U_{(\vec{j},\vec{j}+\vec{e}_I)}^I A_{\mu (\vec{j}+\vec{e}_I)} \,.
\label{eq:DU}
\end{equation}
Following the terminology in
lattice gauge theory,
we may define the lattice spacing in the $I$-th direction as
\begin{equation}
a_I := \frac{1}{g f_I} \qquad (I=1,2,\cdots,d)
\,.
\label{eq:aI}
\end{equation}
We may also define
the compactification radius of the $I$-th direction as 
\begin{equation}
2\pi L_I := N_I a_I = \frac{N_I}{g f_I} \qquad (I=1,2,\cdots,d)\,.
\label{eq:LI}
\end{equation}

The
mass-square matrix of the
gauge fields
in the vacuum
$U_{(\vec{j},\vec{j}+\vec{e}_I)}^I = 1$ can be read off
from the action \eqref{eq:decS4pd}:
\begin{equation}
M_{g}^2 
:= \sum_{I=1}^d M_{g\, I}^2
:= g^2 \sum_{I=1}^d f_I^2
\,\,
\mathbbm{1}_{N_1}
\otimes \cdots
\otimes
\mathbbm{1}_{N_{I-1}}
\otimes
K_{N_I}
\otimes
\mathbbm{1}_{N_{I+1}}
\otimes \cdots
\mathbbm{1}_{N_d}
\,,
\label{eq:Mg2}
\end{equation}
where 
$\mathbbm{1}_{N_J}$ denotes 
the $N_J \times N_J$ identity matrix
and 
$K_{N_I}$ is the $N_I \times N_I$ matrix
given as
\begin{equation}
K_{N_I}:=
\left(
\begin{array}{cccccc}
2      & -1 & 0 & 0 & \cdots & -1\\
-1     & 2 & -1 & 0 & \cdots & 0\\
0 & -1 & 2 & -1 & \cdots & 0\\
\vdots &   &   & \ddots   & & \vdots\\
0   & \cdots &  &   & 2 & -1\\
-1 & \cdots & & & -1 & 2
\end{array}
\right)\,.
\label{eq:matK}
\end{equation}
The mass-square eigenvalues can be obtained
using
the Discrete Fourier Transform (DFT):
\begin{equation}
A_{(\vec{j},\vec{j}+\vec{e}_I)}^I
=
\frac{1}{\prod_{{J}=1}^d N_{J}^{{1}/{2}}}
\sum_{\vec{n}}
\tilde{A}_{(\vec{n})}^I
e^{i \sum_{K=1}^{d} \frac{2\pi n_K j_K}{N_K}}\,.
\label{eq:ADFTHD}
\end{equation}
Our convention for DFT is given in Appendix~\ref{App:DFT}.
In \eqref{eq:ADFTHD}, 
the sum over $\vec{n}$
follows our convention
\eqref{eq:fnodd} or \eqref{eq:fneven} 
for each component $n_I$.
The mass-square eigenvalues $M_{g\, I\, (\vec{n})}^2$
can be parametrized by the discrete Fourier mode $\vec{n}$
and given as
\begin{equation}
M_{g\, I\, (\vec{n})}^2
=
4 g^2 f_I^2 
\sin^2 
\left( 
\frac{\pi n_I}{N_I}
\right)
=
\left(
\frac{2}{a_I}
\right)^2
\sin^2 
\left( 
\frac{\pi n_I}{N_I}
\right)
\,.
\label{eq:MgI2ev}
\end{equation}
The product gauge group 
$\prod_{\vec{j}} U(1)_{(\vec{j})}$
is spontaneously broken to the diagonal $U(1)$
which corresponds to the zero-mode
$\vec{n} = \vec{0}$.
From \eqref{eq:MgI2ev}
and using \eqref{eq:LI}, we observe that
in large $N_I$ limit
the mass spectrum approaches
the Kaluza-Klein (KK) mass spectrum of
the ordinary $d$-dimensional torus with 
the radius of the $I$-th direction being $L_I$.
We will use the same terminology in
continuous extra dimensions
for a corresponding quantity 
in (de)constructed extra dimensions
when the correspondence is obvious
(e.g.~KK scale).

The second line in \eqref{eq:decS4pd}
corresponds to the Wilson plaquette action
in the language of lattice gauge theory.
A natural magnitude of 
the real coupling constant
$\beta_{IJ}$ is of order one,
as can be estimated from
the dimensional analysis
with symmetry considerations
\cite{ArkaniHamed:2001nc,Furuuchi:2011px}.
With $\beta_{IJ}$ of order one,
the extra-dimensional components
of the gauge field
acquire masses of the order of the KK scale
in the vacuum
except for the zero-modes,
like the space-time components 
of the gauge field (see \eqref{eq:MgI2ev}).

The matter Lagrangian density ${\cal L}_{matter}$
in \eqref{eq:decS4pd}
is a sum of
Lagrangian densities of
charged matter fields.
For simplicity, we consider
scalar fields 
$\chi_{(\vec{j})}^q$
which has a charge $q$ under 
the $U(1)_{(\vec{j})}$ 
gauge group.
The charge $q$ of 
the scalar field $\chi_{\vec{j}}^q$
is the same for all $\vec{j}$
to respect the symmetry under 
the discrete translation \eqref{eq:DT}, 
like the gauge coupling $g$.
The $U(1)_{(\vec{j})}$ gauge group is compact
and thus the charges are quantized.
We normalize the gauge coupling $g$
so that all charges in the model are integers.
The Lagrangian density of the 
charged scalar field with charge $q$ is given by
\begin{align}
{\cal L}_{q}
=&
D_\mu \chi_{(\vec{j})}^{q\,\dagger}
D^\mu \chi_{(\vec{j})}^q
-
m^2 \chi_{(\vec{j})}^{q\,\dagger}\chi_{(\vec{j})}^q
\nn\\
&-
\sum_{I=1}^d
\Biggl[
\gamma_I^q f_I^2
\left(
\left(U^I_{(\vec{j},\vec{j}+\vec{e}_I)}\right)^q \chi_{(\vec{j}+\vec{e}_I)}^q
- \chi_{(\vec{j})}^q \right)^{\dagger}
\left(
\left(U^I_{(\vec{j},\vec{j}+\vec{e}_I)}\right)^q \chi_{(\vec{j}+\vec{e}_I)}^q
- \chi_{(\vec{j})}^q \right)
\Biggr]
\,,
\label{eq:Lmatterq}
\end{align}
where the covariant derivative 
for the charged matter with charge $q$
is given as
\begin{equation}
D_\mu \chi_{(\vec{j})}^q
=
\pa_\mu \chi_{(\vec{j})}^q - i g q A_{\mu (\vec{j})} \chi_{(\vec{j})}^q
\,.
\label{eq:covq}
\end{equation}

We will eventually be interested
in the zero-mode of 
the first component of the 
extra-dimensional components of the
gauge field,
which will play the role of the inflaton:
\begin{equation}
\phi
:=
\tilde{A}_{(\vec{0})}^{1}
=
\frac{1}{\prod_{{I}=1}^d N_{I}^{1/2}}
\sum_{\vec{j}}
A_{(\vec{j},\vec{j}+\vec{e}_{1})}^{1}\,.
\label{eq:A0d}
\end{equation}
As shown in Appendix~\ref{app:Voneloop}
(see \eqref{eq:AppVq}),
each 
massless\footnote{As explained in Appendix~\ref{app:Voneloop},
when estimating the one-loop effective potential,
we will treat fields whose mass is far below the KK energy scale
as massless, while we will drop the contributions from
fields whose mass is above the KK energy scale.}
charged scalar field
contributes to the effective potential
of the zero-mode $\phi$ 
at one-loop level as
\begin{equation}
V^q(\phi)
=
-
\Lambda^4
\cos 
\left[
\frac{q \phi}{F}
\right]\,,
\label{eq:Vq}
\end{equation}
where
\begin{align}
\Lambda^4
&=
\frac{1}{2(4\pi)^2}
\frac{2^d}{\pi^{d/2}}
\Gamma \left( 2 + \frac{d}{2} \right)
(2\pi L_1)(2\pi L)^{d-1}
\left(
\frac{4}{(2\pi L_1)^2}
\right)^{2+\frac{d}{2}}
\nn\\
&=
\frac{1}{2(4\pi)^2}
\frac{2^d}{\pi^{{d}/{2}}}
\Gamma\left( 2 + \frac{d}{2} \right)
N_1 N^{d-1} \left( \frac{2}{N_1} \right)^{4+d}
(gf)^4 
\,,
\label{eq:Lambda4}
\end{align}
and
\begin{equation}
F 
=
\frac{N^{\frac{d-1}{2}}f}{N_1^{\frac{1}{2}}} \,.
\label{eq:Ff}
\end{equation}
In the above, we have set 
\begin{align}
f_I &= f \qquad (\mbox{for all $I$})\,,
\label{eq:allf}
\\
N_I &= N \qquad (\mbox{for all $I \ne 1$})\,,
\label{eq:NIN}
\\
N &\gg N_1
\label{eq:NN1}
\,.
\end{align}
From \eqref{eq:LI},
the simplifying assumptions
\eqref{eq:allf} and \eqref{eq:NIN}
make all $L_I$ except for $I=1$ equal, 
which we call $L$: $L_I = L$ for all $I \ne 1$.
Together with the simplifying assumptions
\eqref{eq:allf} and \eqref{eq:NIN},
the condition \eqref{eq:NN1}
can be used to make 
the low energy effective potential such that
$\phi =\tilde{A}_{(\vec{0})}^1$ direction
satisfies the slow-roll condition
while $\tilde{A}_{(\vec{0})}^I$ directions ($I \ne 1$) do not.
Then the model is described as a single-field inflation.

The low energy effective action
which is
appropriate below the 
KK-energy scale $1/L$ is given as
\begin{align}
S_{4}
=&
\int d^4x
\Biggl[
\frac{1}{2} \pa_\mu \phi \pa^\mu \phi
- V(\phi)
\nn\\
&
+
\sum_{\mathrm{charged\,\,matter}}
\sum_{n_1}
\left\{
D_\mu \tilde{\chi}^{q\dagger}_{(n_1)}
D^\mu \tilde{\chi}^{q}_{(n_1)}
-
\tilde{\chi}^{q\dagger}_{(n_1)}
M_{n_1}^2 (q,\phi)
\tilde{\chi}^{q}_{(n_1)}
\right\}
\Biggr]\,.
\label{eq:LEEFT}
\end{align}
%
Here,
the inflaton potential
is given as a sum of 
the contributions \eqref{eq:Vq}
from massless charged scalar fields:
\begin{equation}
V(\phi)
=
C'+\sum_{q} M_q V^q(\phi)\,,
\label{eq:Vinf}
\end{equation}
where $M_q$ is the number of the
massless scalar fields with charge $q$
and $C'$ is the constant.

In the covariant derivative of 
the charged scalar in 
the low energy effective action
\eqref{eq:LEEFT}, 
only the zero-mode of the gauge field 
$A_{\mu(\vec{j})}$ appears:
\begin{equation}
D_\mu \tilde{\chi}^{q}_{(n_1)}
=
\pa_\mu \tilde{\chi}^{q}_{(n_1)}
-
i g_4 q \tilde{A}_{\mu (\vec{0})}
\tilde{\chi}^{q}_{(n_1)}\,,
\label{eq:LEcov}
\end{equation}
where
\begin{equation}
g_4 
:=
\frac{g}{N_1^{\frac{1}{2}}N^{\frac{d-1}{2}}}
\,,
\label{eq:g4}
\end{equation}
is the effective gauge coupling
for the unbroken diagonal $U(1)$ gauge group.
The field $\tilde{\chi}^{q}_{(n_1)}$ is the zero-mode
in the $I\ne 1$ directions, i.e. 
the discrete Fourier mode
$\tilde{\chi}^{q}_{(\vec{n})}$
with $n_I=0$ for $I\ne 1$.
However, the mass of the field  
$\tilde{\chi}^{q}_{(n_1)}$
depends on the expectation value of the inflaton:
\begin{equation}
M_{n_1}^2 (q,\phi) 
=
m^2
+
4 \gamma_1^q f^2
\sin^2
\left(
\frac{q \phi + 2 \pi n_1 F }{2 F N_1}
\right)^2
\,.
\label{eq:Mq}
\end{equation}
As can be seen from \eqref{eq:Mq}, 
which mode number $n_1$ gives the lightest mode
depends on the expectation value of the inflaton field.
Therefore, 
we kept all the discrete Fourier modes labeled by $n_1$.
The inflaton dependent mass \eqref{eq:Mq}
can have interesting consequences
in inflation
\cite{Furuuchi:2015foh,Furuuchi:2020klq},
which we examine in Sec.~\ref{subsec:PP}.

\section{Comparison of explicit models
with CMB observations}\label{sec:obs}

\subsection{Multi-natural inflation 
from (de)construction}\label{subsec:MN}
The simplest
natural inflation model \cite{Freese:1990rb}
is described by a single sinusoidal inflaton potential:
\begin{equation}
V(\phi) 
= 
\frac{V_0}{2}
\left(
1 - \cos \frac{\phi}{F}
\right)\,.
\label{eq:vanipot}
\end{equation}
The single sinusoidal inflaton potential
is not favored by the latest CMB anisotropy data
\cite{Akrami:2018odb}.
However, simple modifications
to the single sinusoidal potential 
may improve the fit to the observational data.
Here, we choose
multi-natural inflation model \cite{Czerny:2014wza}
as such a simple modification 
with an improved fit to the observational data.
The inflaton potential in this model is given by
two sinusoidal potentials with different periodicities:
\begin{equation}
V(\phi)
=
C'
- 
{\Lambda'}^4 
\left[
\cos \left( \frac{\phi}{F} \right)
+
B
\cos \left( \frac{\phi}{A F} + \theta \right)
\right]
\,,
\label{eq:MNpot}
\end{equation}
where $C'$, $\Lambda'$, $A$, $B$ and $\theta$ are
constant parameters.
The (de)construction of extra-natural inflation
we developed in the previous section
provides a microscopic theory
of the multi-natural inflation model.
In terms of the (de)construction model parameters,
the potential \eqref{eq:MNpot} is
parametrized as (see \eqref{eq:Vinf})
\begin{equation}
V(\phi)
=
M_1 \Lambda^4 
\left[
C 
- 
\cos \left( \frac{\phi}{F} \right)
-
B_q
\cos \left( \frac{q \phi}{F} + \theta \right)
\right]
\,.
\label{eq:MNpotDec}
\end{equation}
Comparing \eqref{eq:MNpot} and \eqref{eq:MNpotDec},
we read off the relation between the parameters
of multi-natural inflation and 
those in the microscopic (de)construction model:
${\Lambda'}^4 = M_1 \Lambda^4$,
$A = 1/q$, $B=B_q$ and $C' = M_1 \Lambda^4 C$.
In the (de)construction model,
$M_1$ is identified with 
the number of the massless scalar fields with charge one,
and $B_q$ is given by
\begin{equation}
B_q := \frac{M_q}{M_1}\,,
\label{eq:Bq}
\end{equation}
where
$M_q$ is the number of the massless scalar fields with charge $q$,
see \eqref{eq:Vinf}.
$\Lambda$ is given by 
the (de)construction model parameters
as in \eqref{eq:Lambda4},
while $F$ is given as in \eqref{eq:Ff}.
We will adjust the constant $C$ so that
the value of the potential at its minimum is zero.
This fine-tuning is the usual cosmological constant problem
which we will not address in this article.

While it would be possible 
to construct a microscopic (de)construction model 
which gives rise to non-zero $\theta$ in 
\eqref{eq:MNpotDec},
such a model would need an additional mechanism
to explain it.%
\footnote{
For example, 
non-zero $\theta$ can arise from
an expectation value of an additional gauge field
in the (de)constructed extra dimensions
coupled to the charged scalar fields.
In order for the model to have a non-zero value of $\theta$,
the model should be such that
the corresponding gauge field has desired expectation value.}
For simplicity, in this article
we only consider models in which $\theta$ is zero.
Setting $\theta =0$ fixes the constant $C$
from the requirement that
the value of the potential 
at its minimum is zero:
\begin{equation}
C = \left( B_q + 1 \right) \,,
\label{eq:Cfixed}
\end{equation}
so that
\begin{equation}
V(\phi)
=
M_1 \Lambda^4 
\left[
\left(
1
- 
\cos \left( \frac{\phi}{F} \right)
\right)
+
B_q 
\left(
1
- 
\cos \left( \frac{q \phi}{F} \right)
\right)
\right]
\,.
\label{eq:MNpotThetaZero}
\end{equation}
From the inflaton potential
\eqref{eq:MNpotThetaZero},
the slow-roll parameters are obtained as
\begin{align}
\epsilon(\phi) 
&=
\frac{1}{2}
\left(\frac{V'}{V}\right)^2
=
\frac{1}{2F^2}
\left(
\frac{\sin \left(\frac{\phi}{F}\right)
      +
			B_q q \sin \left(\frac{q \phi}{F} \right)%
			}{%
			\left(
			1 - \cos \left(\frac{\phi}{F} \right)
			\right) 
      +
			B_q 
			\left( 
			1 - \cos \left(\frac{q \phi }{F} \right) 
			\right)%
			}
\right)^2 
	\,,
\label{eq:epsilon}\\
\eta(\phi) 
&= 
\frac{V''}{V} 
= 
-
\frac{%
   \cos \left(\frac{\phi }{F}\right) 
	 +
	 B_q q^2 \cos \left(\frac{q \phi}{F}\right)
	  }{%
	 F^2
   \left( 
	 \left( 1 - \cos \left(\frac{\phi }{F}\right) \right)
	 +
	 B_q \left( 1 - \cos \left(\frac{q \phi}{F} \right) \right)
	 \right)%
	  } \,.
   \label{eq:eta}
\end{align}
Here and below, we work in
the Planck units $M_P=1$.
We will use the subscript 
$end$ to indicate that the value is 
at the time when inflation ends.
More explicitly, we define the 
end of inflation as the time when
the slow-roll condition breaks down:
\begin{equation}
\epsilon(\phi_{end}) = 1 \,.
\label{eq:iend}
\end{equation}
When
$|q \phi_{end} | \ll F$,
the Taylor expansion of $\epsilon(\phi_{end})$
gives
\begin{equation}
\phi_{end} \simeq \sqrt{2} \simeq 1.4 \,.
\label{eq:phiendlargeF}
\end{equation}
In the examples we will study,
$F$ is large enough to justify the approximation
$\phi_{end} =1.4$.
Hence we will use this value for $\phi_{end}$ below.

The number of e-folds in slow-roll inflation is given by
\begin{equation}
    {\cal N}_\ast 
		= 
		\int_{\phi_{end}}^{\phi_\ast} d\phi\left(\frac{V}{V'}\right) \,.
    \label{eq:efolds}
\end{equation}
Here and below, we use the subscript $\ast$ to indicate 
that it is the value when the pivot scale exited the horizon. 
Following the Planck 2018 results \cite{Akrami:2018odb},
we chose the pivot scale to be $0.002$~Mpc$^{-1}$.
The inflaton field value $\phi_\ast$ 
when the pivot scale exited the horizon
is determined by
setting the number of e-folds ${\cal N}_\ast$ 
in \eqref{eq:efolds}.

The scalar power spectral index $n_s$ and 
tensor-to-scalar ratio $r$ are given by 
\begin{align}
    n_s &= 1 - 6\epsilon_\ast + 2\eta_\ast \,,
    \label{eq:specindex}
\\
    r &= 16\epsilon_\ast \,.
    \label{eq:tensorscalar}
\end{align}

\begin{figure}[H]
    \centering
    \includegraphics[width=11cm]{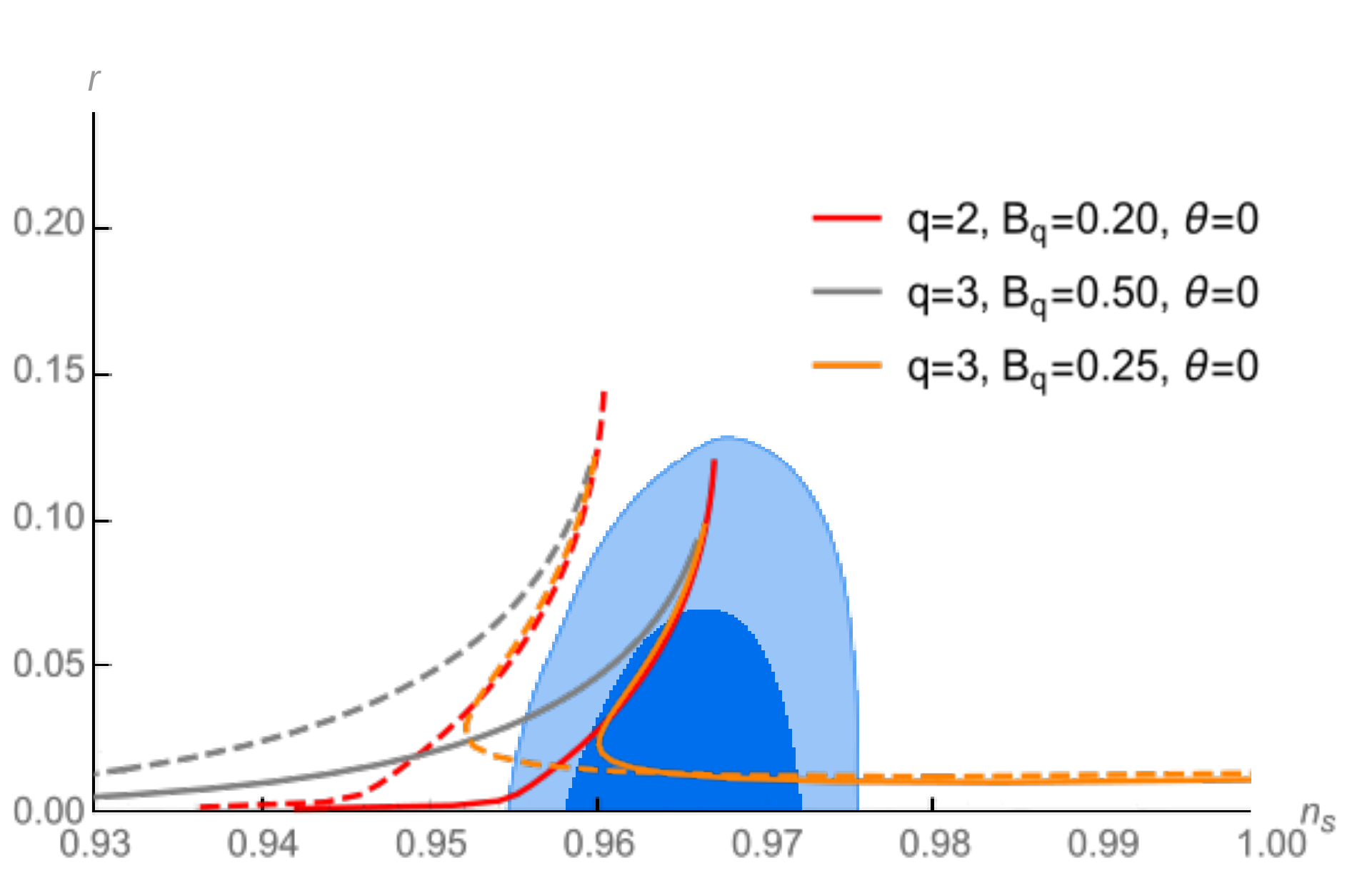}
   \caption{Predicted values of ($n_s,r$) of multi-natural 
   inflation models for three sets of $q$ and $B_q$ values. 
	 The dashed and solid lines 
   correspond to ${\cal N}_\ast = 50$ and ${\cal N}_\ast = 60$,
	 respectively. 
	 These values are compared
	 with the Planck 2018
	 $68\%$ and $95\%$ confidence level regions of ($n_s, r$)
	 (\textit{TT,TE,EE+lowE+lensing})
	 \cite{Akrami:2018odb}.}
    \label{fig:nsvr}
\end{figure}

In Fig.~\ref{fig:nsvr},
we plot the predicted values of 
$n_s$ and $r$ for 
different choices of $B_q$ and $q$,
for
${\cal N}_\ast = 50$ and ${\cal N}_\ast=60$,
for a range of values of $F$.
The predicted values of $n_s$ and $r$
are compared with  
the Planck 2018 results \cite{Akrami:2018odb}.
From Fig.~\ref{fig:nsvr}, 
we observe that the predicted 
values of ($n_s,r$) enter
observationally favored region
for a range of values of $F$.

The scalar power spectrum $P_s$
from slow-roll inflation 
is given by
\begin{equation}
P_s = \frac{H^2(\phi_\ast)}{8\pi^2 \epsilon_\ast} 
= \frac{V(\phi_\ast)}{24\pi^2 \epsilon_\ast} 
= 2.2 \times 10^{-9}\,,
\label{eq:COBE}
\end{equation}
where 
$H$ is the Hubble parameter and
we have used 
the slow-roll approximation
$3H^2(\phi) = V(\phi)$.

In the following, we study
the observational constraints
on models with explicit choices
of $q$, $B_q$ and $M_1$.
We first find the range of $F$ allowed by
the Planck 2018 results 
for a given model. 
Then from the constraint on the parameter $F$, 
we derive the constrains on 
the number of the lattice points $N$ and $N_1$
for a given set of parameters.
We will also examine 
independent constraints
coming from the requirement that
the model should be described
by the low energy action
\eqref{eq:LEEFT} during inflation.
\subsubsection*{The model
\texorpdfstring{${q=2}$}{q}, 
\texorpdfstring{${B_q=0.2}$}{Bq}, 
\texorpdfstring{$M_1=5$}{M1} with 
\texorpdfstring{${\cal N}_\ast =60$}{Nast}}
We first study the model $q=2$, $B_q=0.2$,
$M_1 =5$
with 
${\cal N}_\ast =60$,
which has a good overlap 
with the observationally allowed region
in the $n_s$-$r$ plane Fig.~\ref{fig:nsvr}
for a range of parameters.

\begin{figure}[H]%
    \centering
    {{\includegraphics[width=7cm]{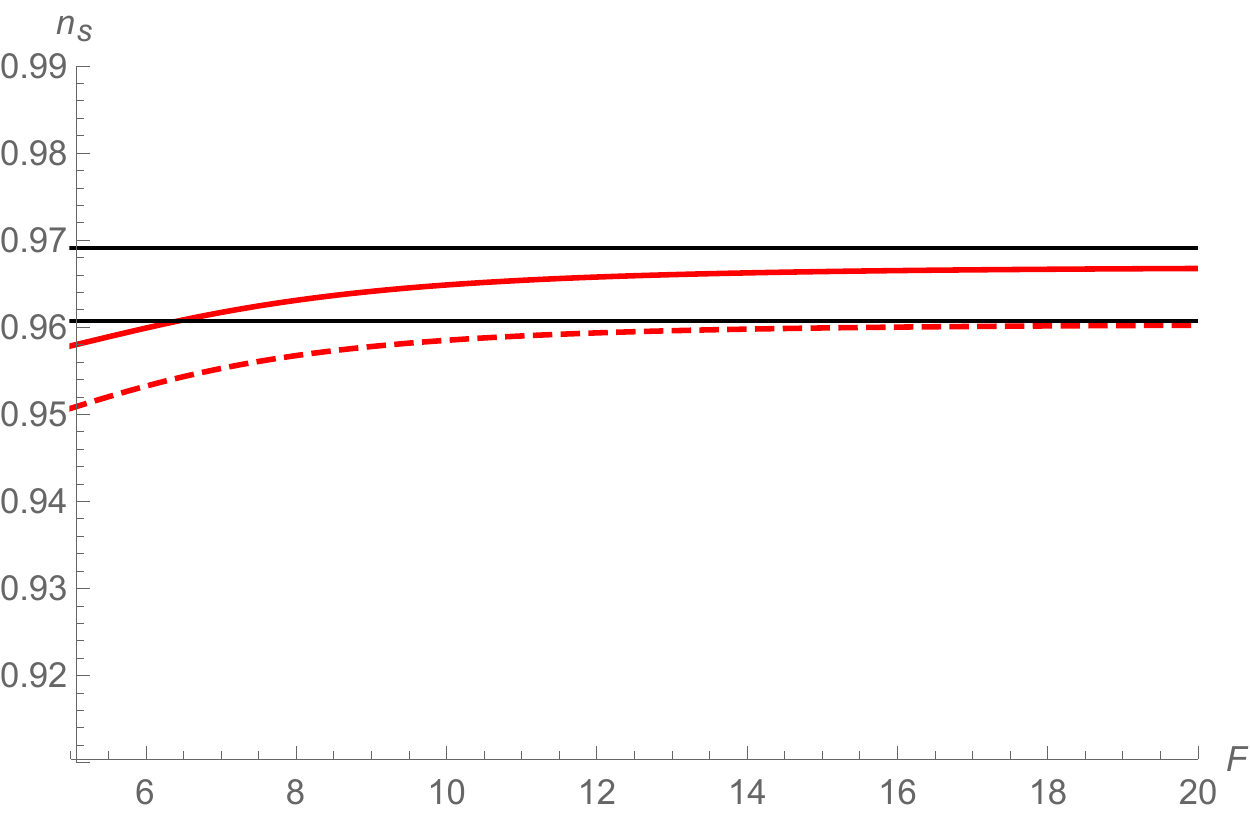} }}%
    \quad
    {{\includegraphics[width=7cm]{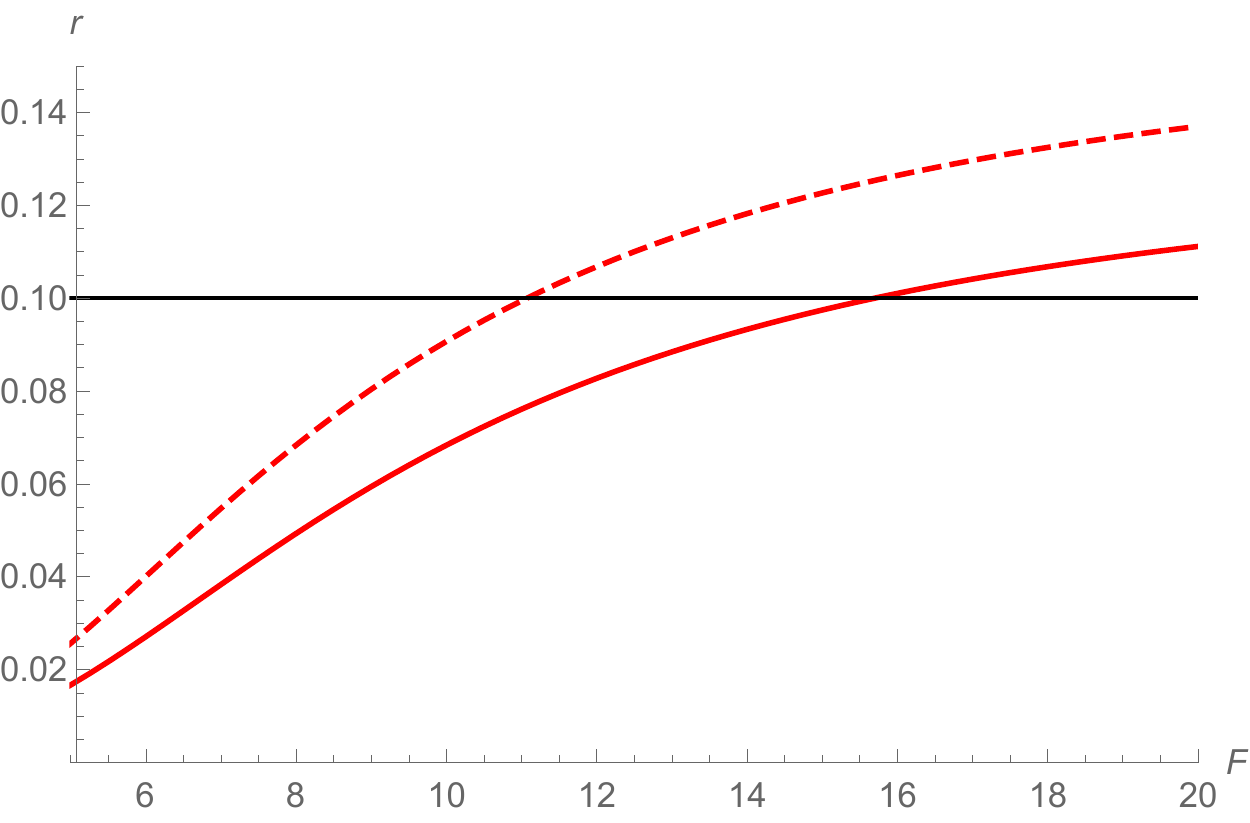} }}%
    \caption{The plot of the power spectral index $n_s$ 
	  (left) and 
    the plot of the tensor-to-scalar ratio $r$ 
		(right) for a range of parameter $F$ 
		for the model $q=2$, $B_q = 0.20$, $M_1=5$ with 
		${\cal N}_\ast =50$ (red dashed line), 
		${\cal N}_\ast =60$ (red bold line). 
		The horizontal lines in the left plot
    show 
		the lower and the upper bounds on $n_s$ with
		$68\%$ confidence level,
		and the horizontal line in the right plot
		shows the upper bound on $r$ 
		with $95\%$ confidence level 
		from the Planck 2018 results
		\cite{Akrami:2018odb}.}%
    \label{fig:ConNsR}%
\end{figure}

\begin{figure}[H]
    \centering
    \includegraphics[width=8cm]{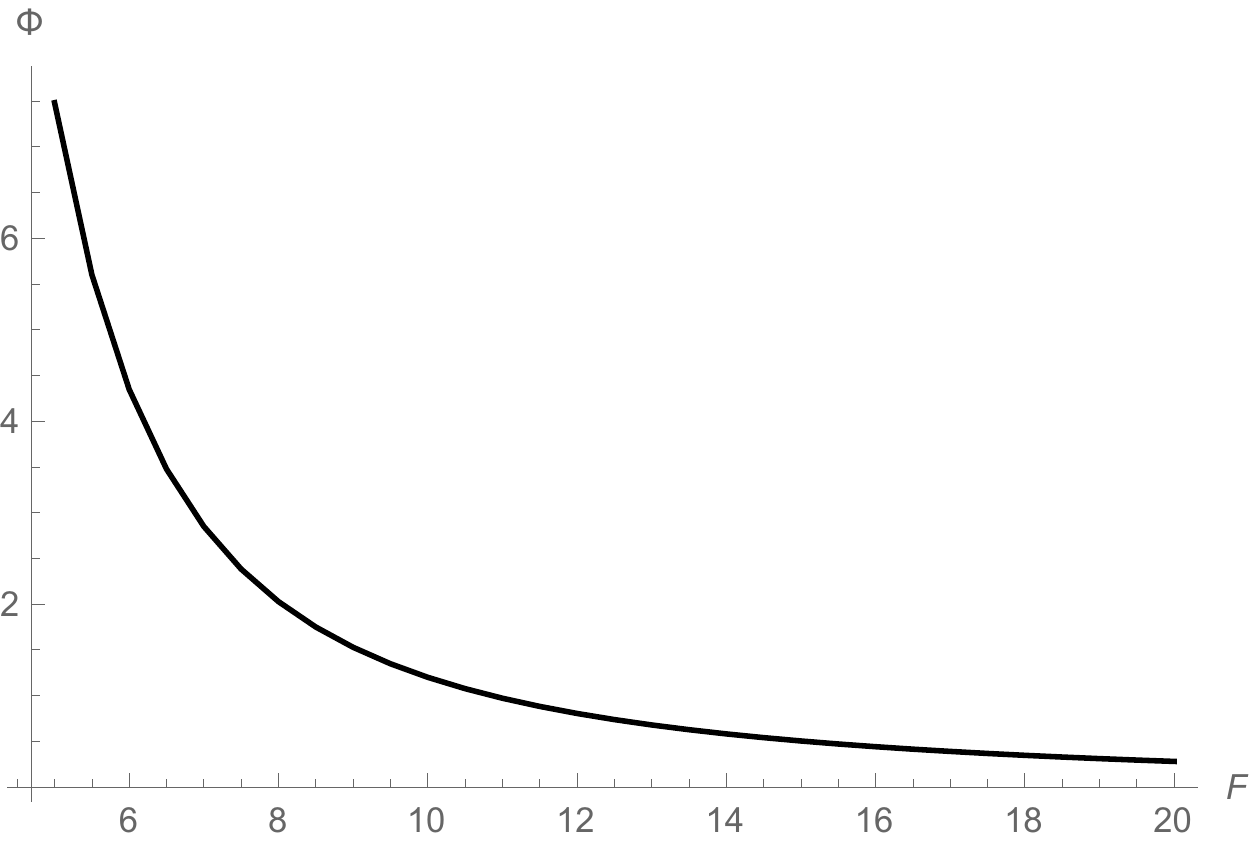}
    \caption{The plot 
		of $\Phi [F]$ given in \eqref{eq:Phi} 
		for a range of the parameter $F$
		for the model $q=2$, $B_q=0.2$, $M_1 =5$
    with ${\cal N}_\ast =60$.}
    \label{fig:Phi}
\end{figure}

On the left in
Fig.~\ref{fig:ConNsR}, 
the spectral index $n_s$ is plotted
for a range of $F$. 
The horizontal lines correspond to
the upper and the lower bounds
on $n_s$ from the Planck 2018 results \cite{Akrami:2018odb} 
with $68\%$ confidence level (\textit{Planck TT,TE,EE+lowE+lensing}):
\begin{equation}
    n_s = 0.9649 \pm 0.0042\, .
    \label{eq:Planck_ns}
\end{equation}
On the right in Fig.~\ref{fig:ConNsR}, 
the tensor-to-scalar ratio $r$ is plotted 
for a range of $F$.
The horizontal line corresponds to 
the upper bound 
given in the Planck 2018 results \cite{Akrami:2018odb} 
with $95\%$ confidence level 
(\textit{Planck TT+lowE+lensing}):
\begin{equation}
    r < 0.10 \, .
\end{equation}
From the $F$-$n_s$ plot on the left of Fig.~\ref{fig:ConNsR}, 
we find the lower bound of $F$ 
for the number of e-folds ${\cal N}_\ast =60$ as
\begin{equation}
    F_{l.b.} = 6.4 \,.
\label{eq:Flb}
\end{equation}
From the $F$-$r$ plot, we find the upper bound
of $F$ for the number of e-folds ${\cal N}_\ast =60$: 
\begin{equation}
    F_{u.b.} = 16 \,.
\label{eq:Fub}
\end{equation}
The lower and the upper bound on $F$,
\eqref{eq:Flb} and \eqref{eq:Fub},
constrain the range of the number 
of the lattice points $N$ and $N_1$.
To see this,
we first 
substitute 
\eqref{eq:MNpotThetaZero}
and 
\eqref{eq:epsilon} 
in 
\eqref{eq:COBE} to obtain
\begin{equation}
    P_s = 
    \frac{%
		M_1\Lambda^4 F^2
		\left(
		\left( 1 - \cos \left(\frac{\phi_\ast}{F}\right) \right) 
		+ 
		B_q 
		\left( 
		1 - \cos \left(\frac{q \phi _\ast}{F}\right) 
		\right)
		\right)^3
	}%
	{
	12 \pi^2 
	\left(
	{\sin \left(\frac{\phi_\ast}{F}\right)}
	+
	{B_q q \sin \left(\frac{q \phi_\ast}{F}\right)}
	\right)^2
	}
	\,.
   \label{eq:PowerSpec}
\end{equation}
For generic values of parameters,
we cannot analytically perform integration 
in \eqref{eq:efolds}
to have explicit functional form
of ${\cal N}_\ast$ as a function of $\phi_\ast$.
However, notice that
from \eqref{eq:efolds} and \eqref{eq:MNpotThetaZero},
once $q$, $B_q$ and ${\cal N}_\ast$ are given,
$\phi_\ast$ only depends on the parameter $F$.
Consequently, from \eqref{eq:PowerSpec}, 
$P_s/M_1 \Lambda^4$ only depends on $F$.
Let us denote this function of $F$ as $\Phi[F]$:
\begin{equation}
\Phi [F] 
:= 
\frac{\left. P_s \right|_{q=2,\, B_q = 0.2,\, {\cal N}_\ast =60}[F]}%
{M_1 \Lambda^4} \,.
\label{eq:Phi}
\end{equation}
In Fig.~\ref{fig:Phi} we plot
$\Phi [F]$ obtained by numerically solving \eqref{eq:efolds}
to obtain $\phi_\ast$,
and putting the obtained value
of $\phi_\ast$ into \eqref{eq:PowerSpec},
for a range of values of $F$.
Since Fig.~\ref{fig:Phi} is numerically evaluated
at each point in $F$,
it may not be easy for the readers to 
read off the value of $\Phi [F]$ for a desired value of $F$.
Therefore, in Appendix~\ref{App:Fit}, 
we provide a fitting function
$\Phi_{fit} [F]$
which reproduces $\Phi [F]$ with around $1\%$ level error or less
for the range of $F$ of interest.

The power spectrum $P_s$ is 
fixed by the COBE normalization \eqref{eq:COBE}:
\begin{equation}
	P_s = M_1 \Lambda^4 \Phi [F]	= 2.2 \times 10^{-9}\,.
    \label{eq:PowerSpecCOBE}
\end{equation}
On the other hand, 
we can write
$\Lambda$ in terms of 
the (de)construction model parameters
as in 
\eqref{eq:Lambda4}: 
\begin{equation}
    \Lambda^4 
		= 
    \frac{1}{2(4\pi)^2}
		\frac{2^d}{\pi^{{d}/{2}}}
		\Gamma \left( 2 + \frac{d}{2}\right)
		N_1 N^{d-1} \left( \frac{2}{N_1} \right)^{4+d}
		(gf)^4 
		\,.
    \label{eq:LambdaN1}
\end{equation}
Substituting \eqref{eq:LambdaN1}
into \eqref{eq:PowerSpecCOBE}
and also using \eqref{eq:Ff} to write
$N_1$ in terms of $f$, $F$ and $N$,
we obtain $N$ for a given 
set of parameters $F$, $g$, $f$, $M_1$ 
and the number of extra dimensions $d$:
\begin{equation}
N
=
\left(
\frac{M_1 \Phi [F]}{2.2 \times 10^{-9}}
\frac{2^{3+2d} \, \Gamma\left( 2 + \frac{d}{2} \right)}{(4\pi)^2 \pi^{d/2}}
\left(
\frac{f^2}{F^2}
\right)^{-(3+d)}
(gf)^{4}
\right)^{\frac{1}{(d-1)(d+2)}}
\,.
\label{eq:Nfunc}
\end{equation}

\begin{figure}[H]
    \centering
    \includegraphics[width=9cm]{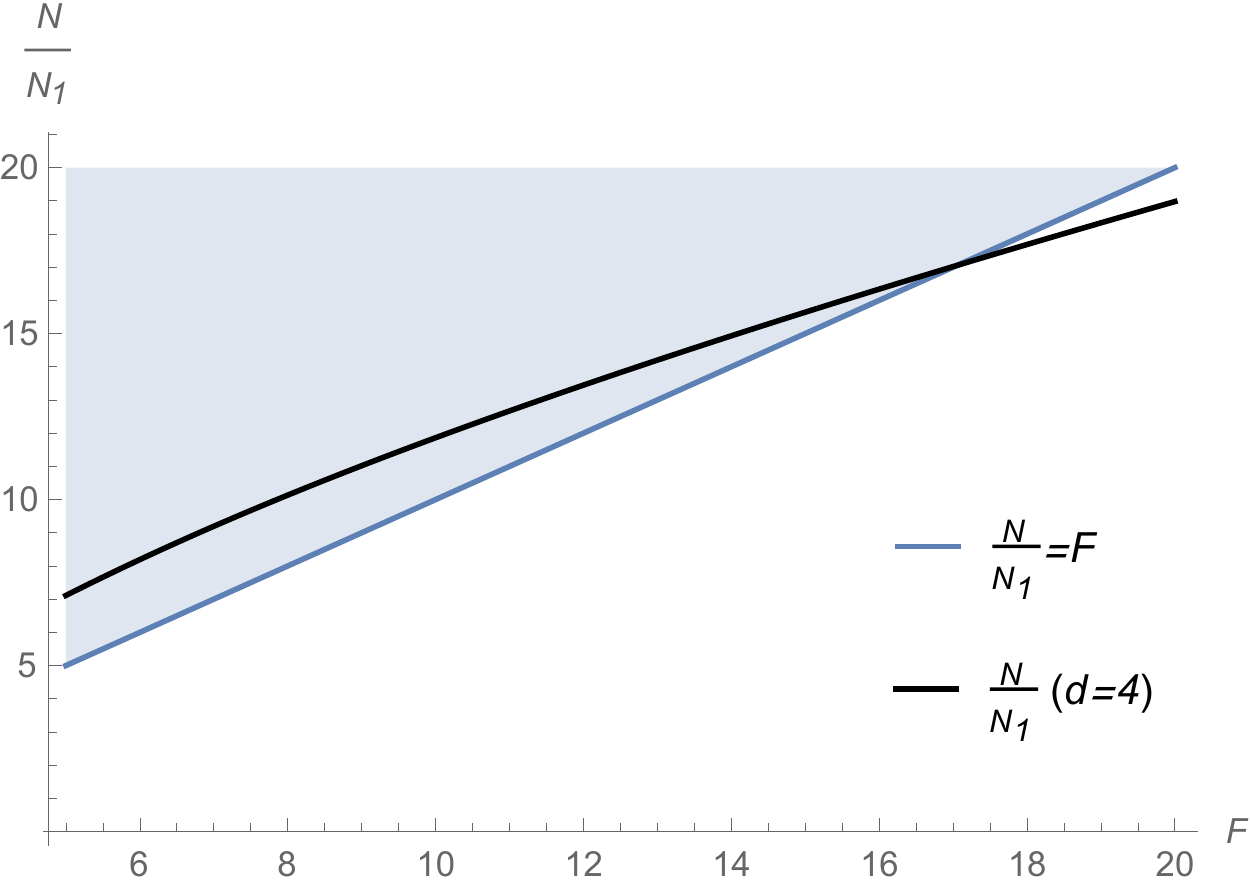}
    \caption{The plot of $N/N_1$ and the constraint on
		it that the model to be described
		as a single-field inflation model \eqref{eq:RNN1F} 
		for the range of $F$ of interest.} 
    \label{fig:NotMFI1}
\end{figure}

From the lower and the upper bound on $F$
\eqref{eq:Flb} and \eqref{eq:Fub},
we obtain constraints on $N$, 
and then through \eqref{eq:Ff}
constraints on $N_1$.
The constraints on $N$ and $N_1$
for the model $q=2$, $B_q=0.2$, $M_1=5$
with ${\cal N}_\ast = 60$,
$g=1.0$, $f=1.0\times 10^{-2}$ 
are summarized in
Table.~\ref{table:NN1range}.
\begin{table}[H]
\begin{center}
  \begin{tabular}{|c|c|c|} \hline
    $d$ & Constraints on $N$ & Constraints on $N_1$ \\ \hline 
    $2$ &  $ 2.6 \times 10^7 \leq N   \leq 1.5 \times 10^8$ 
		    &  
				   $60 \leq N_1 \leq 65$ 
					\\ \hline
	  $3$ &  $ 3.9 \times 10^3 \leq N   \leq 9.4 \times 10^3$ 
		    &  
				 $35 \leq N_1 \leq 37$ 
				 \\ \hline
    $4$ &  $ 2.2 \times 10^2 \leq N   \leq 4.0 \times 10^2 $ 
		    &  $ 24 \leq N_1 \leq 25 $\\ \hline
    $5$ &  
		        $53 \leq N   \leq 83$ 
		    &  $ 19 \leq N_1 \leq 20 $\\ \hline
		$6$ &  
		         $23 \leq N   \leq 33 $ 
		    &  $ N_1 = 16$\\ \hline
  \end{tabular}
	\caption{The constraints on $N$ and $N_1$ derived from the 
	lower and the upper bound on $F$,
	\eqref{eq:Flb} and \eqref{eq:Fub},
	for the model $q=2$, $B_q=0.2$, $M_1=5$ 
	with ${\cal N}_\ast = 60$, $g=1.0$, $f=1.0\times 10^{-2}$.}
	\label{table:NN1range}
	\end{center}
\end{table}
We observe that for $d \geq 5$,
our assumption $N \gg N_1$ 
\eqref{eq:NN1}
may not hold well.
In this case,
the zero-modes of the extra-dimensional components
of the gauge field in other directions
are not too heavier than the inflaton,
and the model may not be described as
a single-field inflation.
Actually, the condition that
the model is described as a single-field inflation,
or more explicitly the condition that 
the low energy effective potential in the direction
of the zero-mode of the $I$-th component ($I\ne 1$)  
of the gauge field does not satisfy the 
slow-roll condition,
can be stated a little bit more precise than \eqref{eq:NN1}.
The periodicity $2\pi F'$ of the zero-mode in $I$-th direction ($I\ne 1$)
is obtained as (see Appendix~\ref{app:Voneloop} eq.\eqref{eq:Vex})
\begin{equation}
F' 
= 
\frac{N_1^{\frac{1}{2}} N^{\frac{d-1}{2}} f}{N}
=
\frac{N_1}{N} F \,.
\label{eq:Fp}
\end{equation}
The slow-roll parameters in the $I$-th direction ($I\ne 1$)
is of the order of $1/{F'}^2$.
Thus the condition that the $I$-th direction ($I\ne 1$)
does not satisfy the slow-roll condition is 
\begin{equation}
\frac{1}{{F'}^2} \gtrsim 1 
\,\,
\Rightarrow
\,\, 
\frac{N}{N_1} \gtrsim F 
\,.
\label{eq:RNN1F}
\end{equation}
Indeed, the condition \eqref{eq:RNN1F}
does not hold for the cases $d=5$ and $d=6$.
In the case $d=4$,
the left hand side and the right hand side
of the inequality in \eqref{eq:RNN1F}
are of the same order and
we may better have a closer look.
In Fig.~\ref{fig:NotMFI1},
$N/N_1$ and $F$ are plotted
for the range of $F$ of interest.
We observe that two lines in the plot
intersect at $F = 17$, which is 
beyond $F_{u.b.}$ \eqref{eq:Fub}.
Therefore, the condition
\eqref{eq:RNN1F} does not give a new constraint
to this model.
Also notice that the inequality in
\eqref{eq:RNN1F} allows 
the both sides to be around the same. 
Thus the region $F > 17$ should not be
excluded immediately by \eqref{eq:RNN1F}.
In fact, we observe from Fig.~\ref{fig:NotMFI1}
that the both sides of \eqref{eq:RNN1F} 
are around the same throughout the range of $F$ of interest.

\begin{figure}[H]
    \centering
    \includegraphics[width=9cm]{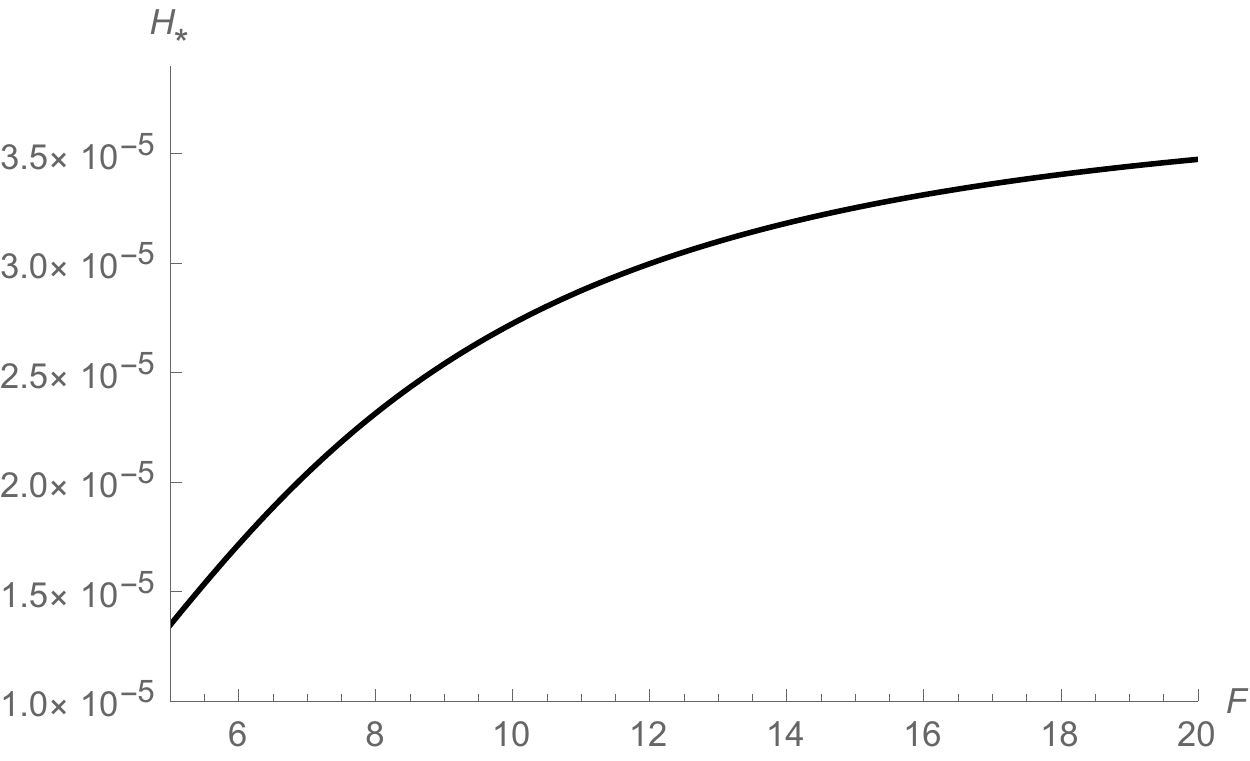}
    \caption{The plot of the  
		Hubble parameter at the pivot scale
		for the range of $F$ of interest
		$F_{l.b.} = 6.4 \leq F \leq F_{u.b.} = 16$.} 
    \label{fig:Hubbleparameter}
\end{figure}

We should also require
the energy scale during inflation
to be lower than the KK energy scale:
\begin{equation}
H \ll 
\frac{1}{L}
= \frac{gf}{N} \,.
\label{eq:HgrtLinv}
\end{equation}
This condition is required for 
the low energy effective action
\eqref{eq:LEEFT}
to be valid during inflation.
In slow-roll inflation
in which the inflaton rolls down
the potential as time elapses, 
the Hubble parameter
at the pivot scale
$H_\ast:= H(\phi_\ast)$
is close to the maximum value
during the observable inflation.
Therefore, we choose $H_\ast$ 
as the representative value of $H$ 
in \eqref{eq:HgrtLinv}.
The value of $H_\ast$ 
is obtained from 
\eqref{eq:COBE}.
For fixed  
$q$, $B_q$ and ${\cal N}_\ast$ which
we chose to be $q=2$, $B_q = 0.2$
and ${\cal N}_\ast = 60$,
$H_\ast$ is a function only on the parameter $F$.
The numerically evaluated values of $H_\ast$
are plotted for the range of $F$ of interest 
in Fig.~\ref{fig:Hubbleparameter}.
Like we did for $\Phi [F]$, 
we also provide
a fitting function for the square of 
the Hubble parameter in Appendix~\ref{App:Fit}.
From Fig.~\ref{fig:Hubbleparameter} 
we observe that $H_\ast$ is of the order of $10^{-5}$.
Putting $g \sim 1$ and $f \sim 10^{-2}$,
\eqref{eq:HgrtLinv} gives $ N \ll 10^3$.
Comparing this constraint with
Table.~\ref{table:NN1range},
the cases $d \leq 3$ are excluded for
these values of parameters $g$ and $f$,
while the $d \geq 5$ cases are safely
in the allowed region.
In the case $d=4$,
the allowed values of $N$ in Table.~\ref{table:NN1range}
are comparable 
with the boundary of the constraint \eqref{eq:HgrtLinv}
in the range of $F$ of interest,
so we should have a closer look.

\begin{figure}[H]
    \centering
    \includegraphics[width=9cm]{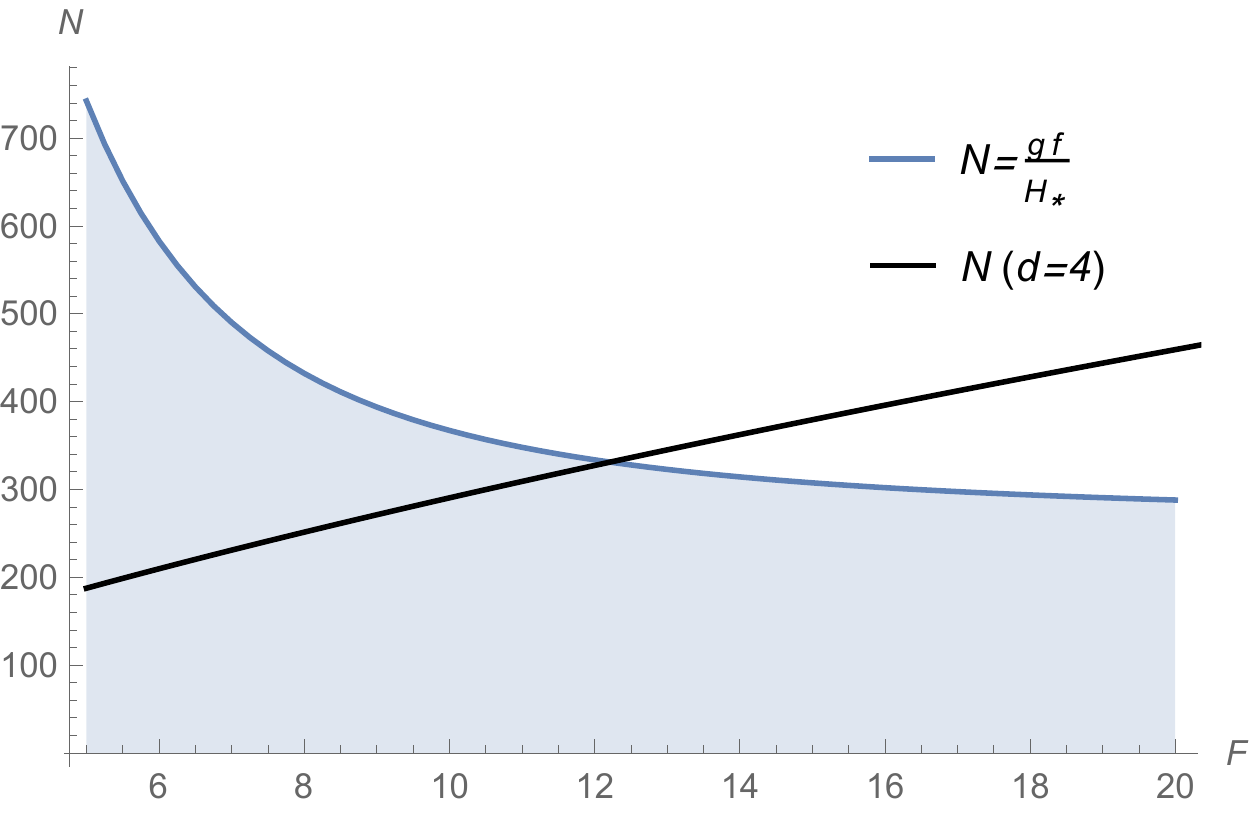}
    \caption{The constraint on $N$ placed by 
		\eqref{eq:HgrtLinv} 
		for the range of $F$ of interest
		$F_{l.b.} = 6.4 \leq F \leq F_{u.b.} = 16$
		for the model $q=2$, $B_q=0.2$, $M_1=5$
    with ${\cal N}_\ast = 60$.
		The parameters $g$ and $f$ are fixed as 
		$g=1.0$ and $f=1.0 \times 10^{-2}$ here.  
		$N$ for the case $d=4$ in the same model 
		with the same values of the parameters
		is plotted to be compared with the constraint.}
    \label{fig:HcNd4}
\end{figure}

In Fig.~\ref{fig:HcNd4}, we plotted
$N$ for the case $d=4$
and the constraint from \eqref{eq:HgrtLinv}.
We observe that the region
$F > 12$ is excluded
by the constraint 
\eqref{eq:HgrtLinv}
for the case $d=4$,
and as a consequence 
$N$ is restricted as $N \leq 3.3 \times 10^2$.


To summarize the results
of this model with the parameter values
${\cal N}_\ast=60$, 
$g=1.0$ and $f=1.0\times 10^{-2}$,
the cases $d\leq 3$ are excluded by the condition
\eqref{eq:HgrtLinv},
while in the cases $d\geq 5$ 
the condition \eqref{eq:RNN1F} is not satisfied.
In the case $d=4$,
we obtain $F \geq 6.4$ 
(which corresponds to $N \geq 2.2 \times 10^2$ and $N_1 = 25$)
from the lower bound on $n_s$ 
as in \eqref{eq:Flb},
and we obtain $F \leq 12$ 
(which corresponds to $N \leq 3.3 \times 10^2$ and $N_1 = 25$)
from \eqref{eq:HgrtLinv}.
Notice that in this range of the parameter $F$,
the predicted tensor-to-scalar ratio $r$ 
is above $0.01$, as seen in Fig.~\ref{fig:ConNsR}.
Via the Lyth bound \cite{Lyth:1996im}, this means that
the model belongs to large-field inflation models 
which enjoy the trans-Planckian inflaton field excursion.

\subsubsection*{The model
\texorpdfstring{$q=3$}{q}, 
\texorpdfstring{$B_q=0.25$}{Bq}, 
\texorpdfstring{$M_1=4$}{M1} 
with 
\texorpdfstring{${\cal N}_\ast =60$}{Nast}}
Next we study the model $q=3$, $B_q=0.25$,
$M_1 =4$
with 
${\cal N}_\ast =60$.
Since the methodology is the same as in the previous model,
we skip the explanations and only quote the results.

\begin{figure}[H]%
    \centering
    {{\includegraphics[width=7cm]{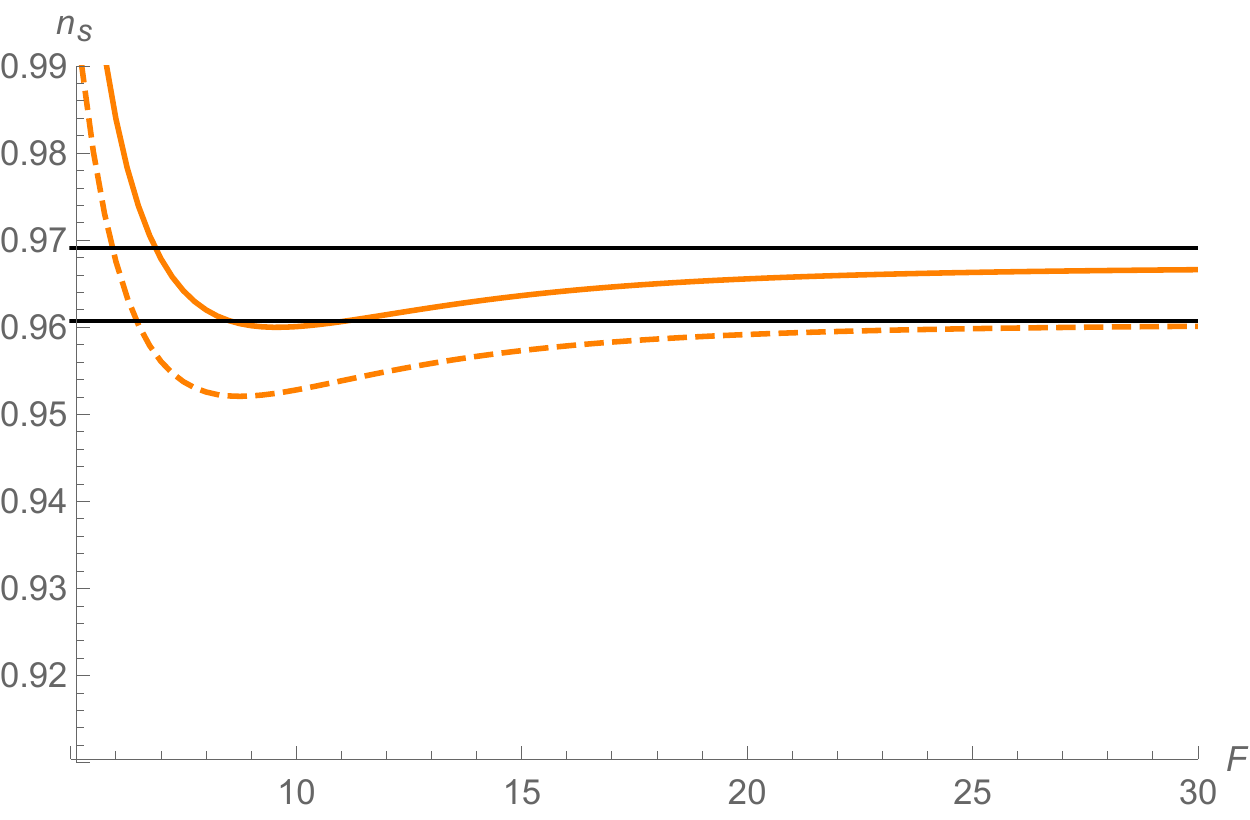} }}%
    \quad
    {{\includegraphics[width=7cm]{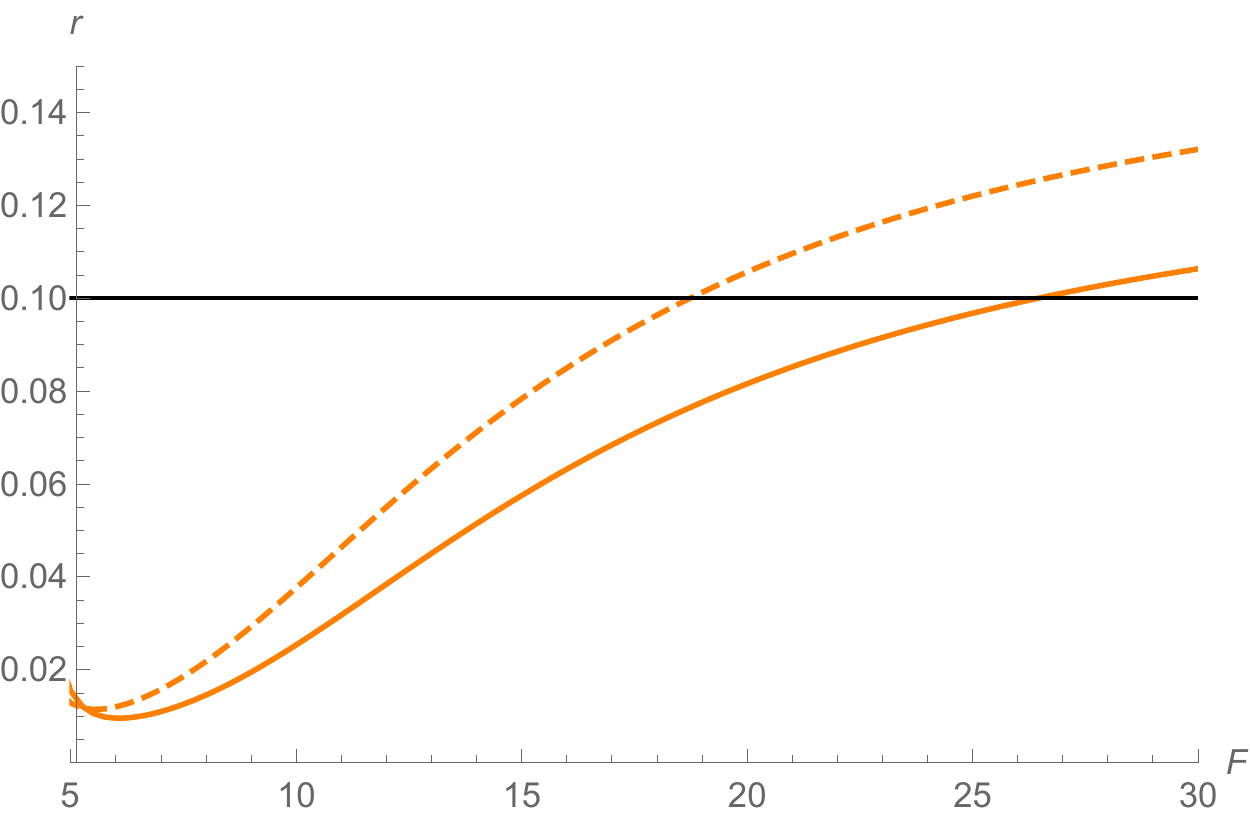} }}%
    \caption{The plot of the power spectral index $n_s$ 
	  (left) and 
    the plot of the tensor-to-scalar ratio $r$ 
		(right) for a range of the parameter $F$
		for the model $q=3$, $B_q = 0.25$, $M_1 =4$ 
		with
		${\cal N}_\ast =50$ (orange dashed line), 
		${\cal N}_\ast =60$ (orange bold line). 
		The horizontal lines in the left plot
    show the lower and the upper bounds on $n_s$ 
		with $68\%$ confidence level,
		and the horizontal line in the right plot
		shows the upper bound on $r$ 
		with $95\%$ confidence level 
		from the Planck 2018 results
		\cite{Akrami:2018odb}.}%
    \label{fig:ConNsR2}%
\end{figure}

From the $F$-$n_s$ plot on the left of 
Fig.~\ref{fig:ConNsR2}, 
we find
the lower bound of $F$ from the upper bound on $n_s$: 
\begin{equation}
    F_{l.b.} = 6.9 \,.
\label{eq:Flb2}
\end{equation}
Here, although in the plot of $n_s$ in 
Fig.~\ref{fig:ConNsR2},
the model prediction slightly comes below
the observational lower bound with $68\%$ confidence level
in the region around $F = 9 \sim 10$,
we did not exclude this region as the 
differences from the lower bound are tiny.

From the $F$-$r$ plot on the right, 
we find the upper bound of $F$: 
\begin{equation}
    F_{u.b.} = 26 \,.
\label{eq:Fub2}
\end{equation}

With a bit of abuse of notation, 
we define the function $\Phi [F]$ 
in the same way as in \eqref{eq:Phi} 
but for the current model:
\begin{equation}
\Phi [F] 
:= 
\frac{\left. P_s \right|_{q=3,\, B_q = 0.25,\, {\cal N}_\ast =60}[F]}%
{M_1 \Lambda^4} \,.
\label{eq:Phi2}
\end{equation}
The numerically evaluated values of $\Phi [F]$
for a range of values of $F$ are
plotted in Fig.~\ref{fig:PhiF2}.
We also provide a fitting function 
$\Phi_{fit} [F]$
in Appendix~\ref{App:Fit}.
\begin{figure}[H]
    \centering
    \includegraphics[width=8cm]{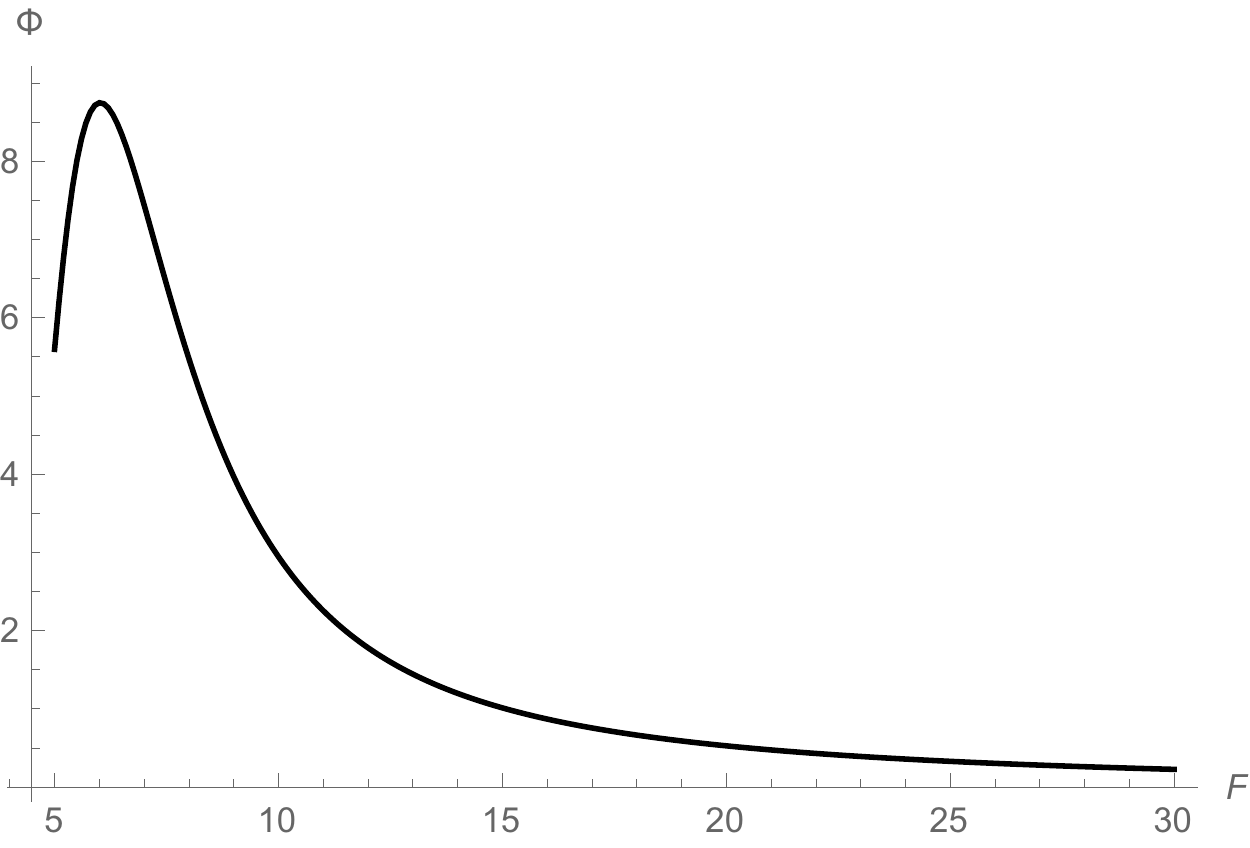}
    \caption{The plot of $\Phi [F]$ given in \eqref{eq:Phi2}
		for a range of $F$ for the model $q=3$, $B_q=0.25$,
    $M_1 =4$ with ${\cal N}_\ast =60$.}
    \label{fig:PhiF2}
\end{figure}

\begin{figure}[H]
    \centering
    \includegraphics[width=9cm]{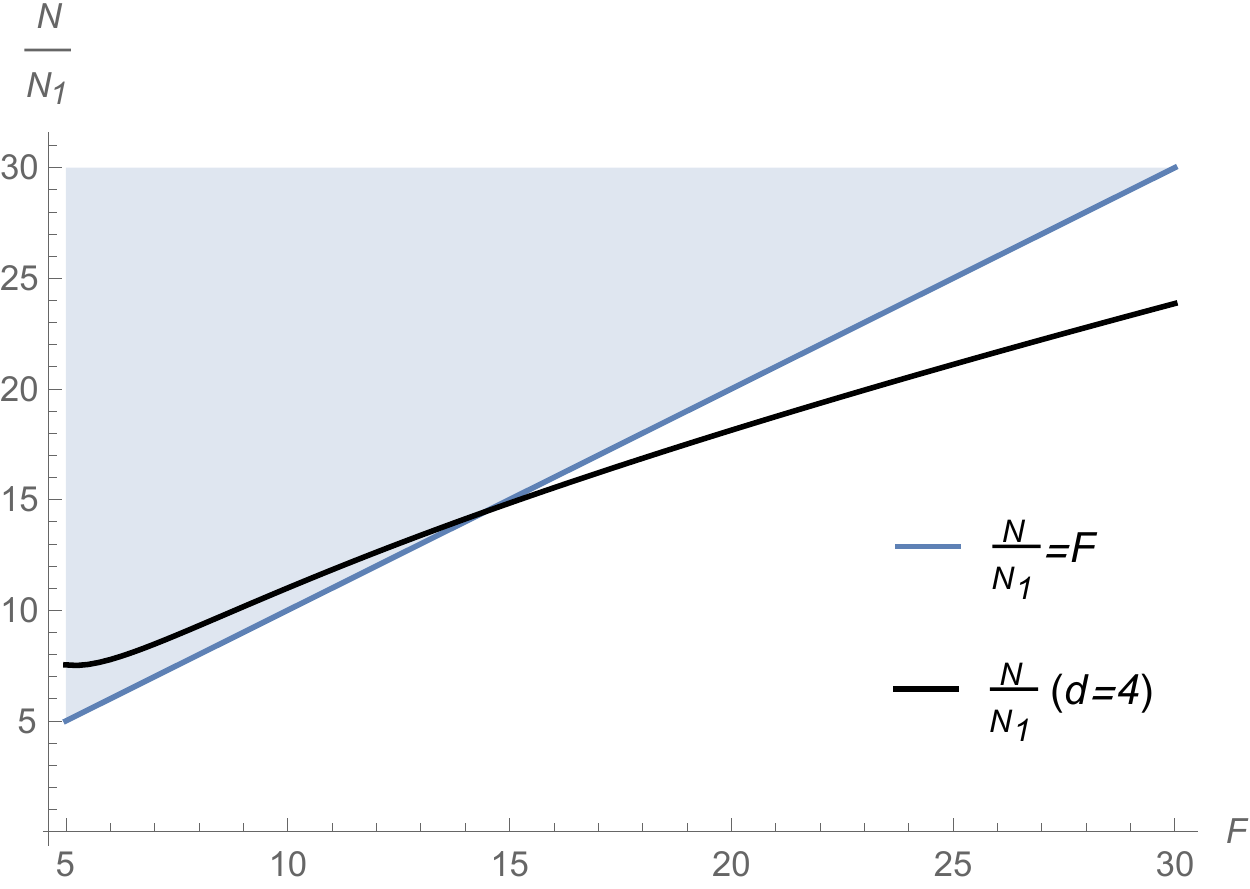}
    \caption{The plot of $N/N_1$ and the constraint on
		it that the model to be described
		as a single-field inflation model \eqref{eq:RNN1F} 
		for the range of $F$ of interest.} 
    \label{fig:NotMFI2}
\end{figure}

Using \eqref{eq:Nfunc},
from the lower and the upper bound on $F$,
\eqref{eq:Flb2} and \eqref{eq:Fub2},
we obtain constraints on $N$, 
and then through \eqref{eq:Ff}
constraints on $N_1$.
The constraints on $N$ and $N_1$
for the model $q=3$, $B_q=0.25$, $M_1=4$
with ${\cal N}_\ast = 60$, 
$g=1.0$, $f=1.0\times 10^{-2}$ 
are summarized in
Table.~\ref{table:NN1range2}.
\begin{table}[H]
\begin{center}
  \begin{tabular}{|c|c|c|} \hline
    $d$ & Constraints on $N$ & Constraints on $N_1$ \\ \hline 
    $2$ &  $ 3.6 \times 10^7 \leq N \leq 4.5 \times 10^8 $ 
		    &  
				 $66 \leq N_1 \leq 76 $ 
				\\ \hline
    $3$ &  $ 4.5 \times 10^3 \leq N \leq 1.6 \times 10^4 $ 
		    &  
				$37 \leq N_1 \leq 42 $ 
				\\ \hline
    $4$ &  $ 2.4 \times 10^2 \leq N   \leq 5.6 \times 10^2 $ 
		    &  
				   $26 \leq N_1 \leq 28$ 
					 \\ \hline
    $5$ &  
		       $57 \leq N   \leq 1.1 \times 10^2 $ 
		    &  
				   $20 \leq N_1 \leq 22$ 
					\\ \hline
		$6$ &  
		       $24 \leq N  \leq 41$ 
		    &  $ 17 \leq N_1 \leq 18 $\\ \hline
  \end{tabular}
	\caption{The constraints on $N$ and $N_1$ derived from the 
	lower and upper bound on $F$, 
	\eqref{eq:Flb2} and \eqref{eq:Fub2},
	for the model $q=3$, $B_q=0.25$, $M_1=4$
	with ${\cal N}_\ast = 60$, $g=1.0$, $f=1.0\times 10^{-2}$.}
	\label{table:NN1range2}
	\end{center}
\end{table}
We observe that for the cases $d \geq 5$,
the condition \eqref{eq:RNN1F}
does not hold for the cases $d=5$ and $d=6$.
In the case $d=4$, 
the left hand side and the right hand side
of the inequality in \eqref{eq:RNN1F}
are of the same order and
we may better have a closer look.
In Fig.~\ref{fig:NotMFI2},
$N/N_1$ and $F$ are plotted
for the range of $F$ of interest.
We observe that two lines in the plot
intersect at $F = 14$.
However, notice that
the inequality in \eqref{eq:RNN1F} 
allows the both sides to be around the same. 
Therefore, we should not rule out
the region $F > 14$ immediately.
In fact, we observe from Fig.~\ref{fig:NotMFI2}
that the both sides of \eqref{eq:RNN1F} 
are around the same throughout the range of $F$ of interest.

\begin{figure}[H]
    \centering
    \includegraphics[width=9cm]{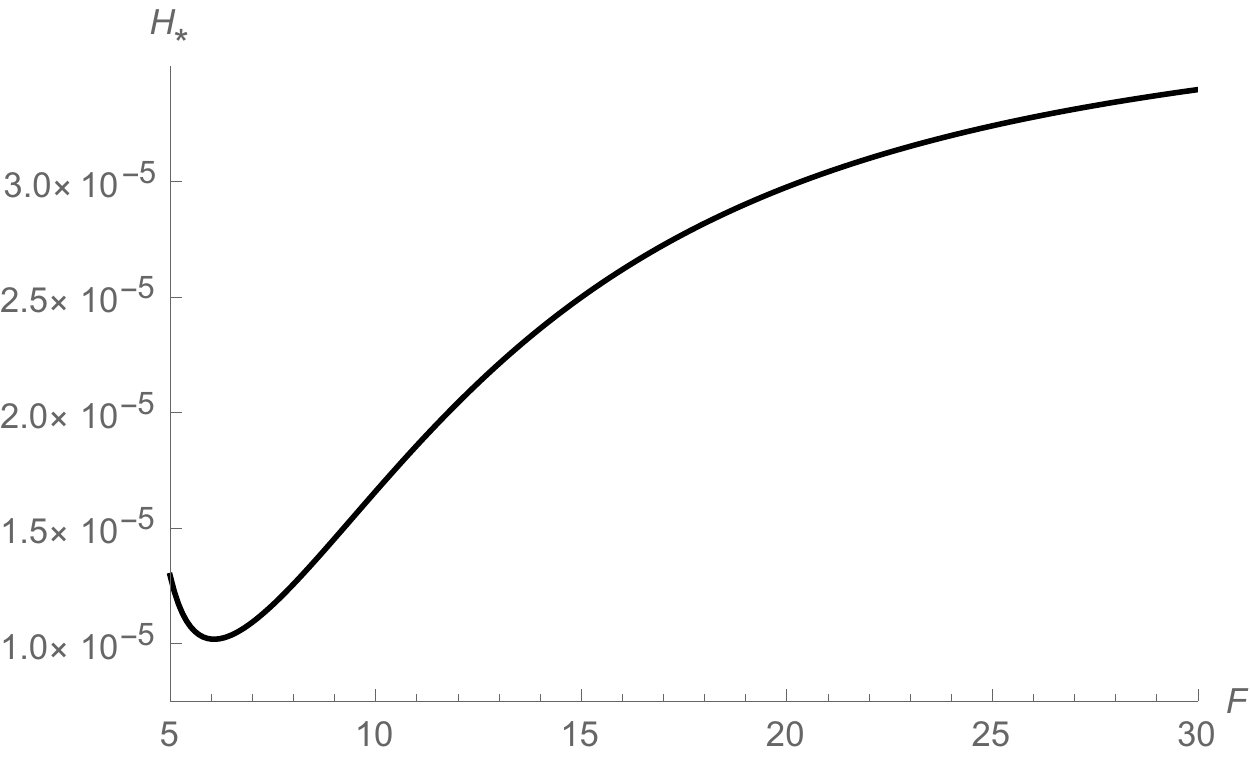}
    \caption{The plot of the  
		Hubble parameter at the pivot scale
		for the model $q=3$, $B_q=0.25$, $M_1 =4$
    with ${\cal N}_\ast =60$
		for the range of $F$ of interest
		$F_{l.b.} = 6.9 \leq F \leq F_{u.b.} = 26$.} 
    \label{fig:Hubbleparameter2}
\end{figure}
\begin{figure}[H]
    \centering
    \includegraphics[width=9cm]{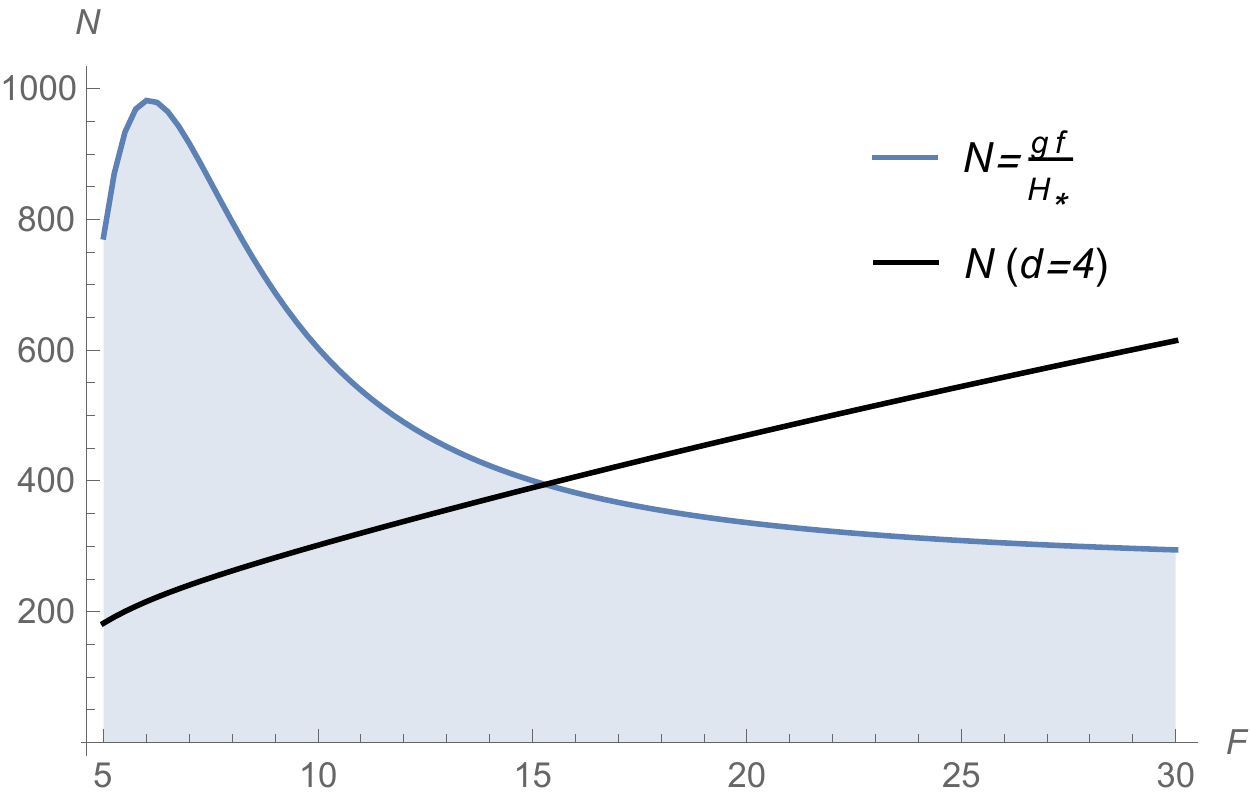}
    \caption{The constraint on $N$ placed by 
		\eqref{eq:HgrtLinv}
		for the model $q=3$, $B_q=0.25$, $M_1 =4$
    with ${\cal N}_\ast =60$
		for the range of $F$ of interest
		$F_{l.b.} = 6.9 \leq F \leq F_{u.b.} = 26$.
		The parameters $g$ and $f$ are fixed as 
		$g=1.0$ and $f=1.0 \times 10^{-2}$ here.  
		$N$ for the case $d=4$ in the same model with 
		the same values of the parameters 
		is plotted to be compared with the constraint.}
    \label{fig:HcNd42}
\end{figure}

We should also 
examine the constraint \eqref{eq:HgrtLinv}.
We plot the numerically evaluated values of
$H_\ast$ for the range of $F$ of interest
in Fig.~\ref{fig:Hubbleparameter2}.
We observe that $H_\ast$ is of the order of $10^{-5}$.
Putting $g \sim 1$ and $f \sim 10^{-2}$,
\eqref{eq:HgrtLinv} gives $ N \ll 10^3$.
Comparing this constraint with Table.~\ref{table:NN1range2},
the cases $d\leq 3$ are excluded for
these values of parameters $g$ and $f$,
while the $d\geq 5$ cases are safely
in the allowed region.
In the case $d=4$,
the allowed values of $N$ in Table.~\ref{table:NN1range2}
are comparable with
the boundary of the constraint \eqref{eq:HgrtLinv}
in the range of $F$ of interest,
so we should have a closer look.
In Fig.~\ref{fig:HcNd42}, we plotted
$N$ for the case $d=4$
and the constraint from \eqref{eq:HgrtLinv}.
We observe that the region
$F > 15$ is excluded
by the constraint 
\eqref{eq:HgrtLinv}
for the case $d=4$,
and as a consequence 
$N$ is restricted as $N \leq 3.9 \times 10^2$
and $N_1$ is restricted as $N_1 \geq 26$.


To summarize the results
of this model 
with parameter values 
${\cal N}_\ast=60$, 
$g=1.0$ and $f=1.0\times 10^{-2}$,
the cases $d \leq 3$ are excluded by the condition
\eqref{eq:HgrtLinv},
while the cases $d \geq 5$, 
the condition \eqref{eq:RNN1F} 
is not satisfied.
In the case $d=4$,
we obtain $F \geq 6.9$ 
(which corresponds to $N \geq 2.4 \times 10^2$ and $N_1 \leq 28$)
from the lower bound on $n_s$ 
as in \eqref{eq:Flb2},
and we obtain $F \leq 15$ 
(which corresponds to $N \leq 3.9 \times 10^2$ and $N_1 \geq 26$)
from \eqref{eq:HgrtLinv}.
Notice that in this range of the parameter $F$,
the predicted tensor-to-scalar ratio $r$ 
is above $0.01$, as seen in Fig.~\ref{fig:ConNsR2}.
Like the previous model,
this means via the Lyth bound \cite{Lyth:1996im} that
the model is a large-field inflation model.

\subsection{Particle production during 
inflation}\label{subsec:PP}

The coupling of the inflaton and 
the charged matter fields
in \eqref{eq:LEEFT}
is of the form investigated in
\cite{Furuuchi:2015foh,Furuuchi:2020klq},
which leads to rapid particle productions
during inflation
and may leave observable 
features in primordial density perturbation
\cite{Kofman:1997yn,Chung:1999ve,Barnaby:2009mc,Barnaby:2009dd,Pearce:2017bdc}.
Below, 
we examine the detectability of 
the primordial features produced
by the rapid particle productions
during inflation.

The mass term
of the charged matter with charge $q$ 
is given as \eqref{eq:Mq}
%
\begin{equation}
\tilde{\chi}^{q\dagger}_{(n_1)} 
M_{n_1}^2 (q,\phi) 
\tilde{\chi}^{q}_{(n_1)}
=
\tilde{\chi}^{q\dagger}_{(n_1)}
4 \gamma_1^q f^2
\sin^2
\left(
\frac{q \phi + 2 \pi n_1 F }{2 F N_1}
\right)^2
\tilde{\chi}^{q}_{(n_1)},
\label{eq:M2chi}
\end{equation}
where we have dropped $m^2$ term in \eqref{eq:Mq},
since only the fields with $m^2 \ll H^2$
will be relevant in the discussions below,
whose effects 
can be well approximated by taking $m^2 = 0$.
Near 
$\phi = - 2 \pi n_1 F/q$,
the inflaton-dependent 
mass term of the matter field
can be approximated as
\begin{equation}
\tilde{\chi}^{q\dagger}_{(n_1)}
M_{n_1}^2 (q,\phi)
\tilde{\chi}^{q}_{(n_1)}
\simeq
\tilde{\chi}^{q\dagger}_{(n_1)}
\frac{\gamma_1^q q^2}{N_1 N^{d-1}}
\left(
\phi + 
\frac{2 \pi n_1 F}{q}
\right)^2
\tilde{\chi}^{q}_{(n_1)}.
\label{eq:M2chiAppr}
\end{equation}
From the analytic result
of \cite{Pearce:2017bdc},
the contribution of the rapid
particle production due to 
the interaction \eqref{eq:M2chiAppr}
to the power spectrum 
$\delta P_s$ 
is given by
\begin{equation}
\delta := 
\frac{\delta P_s}{P_s}
\simeq
2 \times 300 
\left(
\frac{qg}{N_1^{\frac{1}{2}} N^{\frac{d-1}{2}}}
\right)^{7/2}
\,,
\label{eq:deltaPS}
\end{equation}
where the factor $2$ in the right hand side
is from the fact that the complex 
field $\chi^{q}_{(n_I)}$
has two real degrees of freedom.
We also restrict ourselves
to the the universality restoration point
$\gamma_I^q = g^2$ $(I=1,\cdots,d)$,
at which the KK mass spectra of 
the charged scalar fields coincide
with that of the gauge field 
\cite{Furuuchi:2020klq}.
For the detectability of the primordial
feature in the near future, 
we require 
that the amplitude of the feature
to be more than a percent
of the power spectrum: 
%
\begin{equation}
\delta > 0.01 \,.
\label{eq:deltabounds}
\end{equation}
Substituting \eqref{eq:deltaPS}
into \eqref{eq:deltabounds}
and 
the validity of the perturbation theory
$gq \lesssim 1$,
we obtain
\begin{equation}
N_1 N^{d-1} \lesssim 6 \times 10^2 \,.
\label{eq:featureN}
\end{equation}
Using \eqref{eq:Ff},
\eqref{eq:featureN} can be rewritten as
\begin{equation}
N_1^2 \lesssim 6 \times 10^2 \, \frac{f^2}{F^2} \,.
\label{eq:featureN1}
\end{equation}
The EFT described by the action
\eqref{eq:decS4pd}
is valid below $4\pi f$.
It is natural to assume that
this UV cut-off scale is still much below
the Planck scale:
\begin{equation}
4\pi f \ll M_P \,.
\label{eq:fMP}
\end{equation}
On the other hand,
if we restrict ourselves to the large field inflation,
the period $2\pi F$ should be super-Planckian:
\begin{equation}
F \gtrsim M_P \,.
\label{eq:FMP}
\end{equation}
Putting 
\eqref{eq:fMP} and \eqref{eq:FMP} into \eqref{eq:featureN1},
we obtain
\begin{equation}
N_1^2 \ll 4 \,.
\label{eq:N1toosmall}
\end{equation}
The condition \eqref{eq:N1toosmall}
cannot be satisfied with an integer $N_1$.
Thus our (de)constructed model of extra-natural inflation
does not produce detectable primordial feature
under the rather general assumptions on the parameters,
\eqref{eq:fMP} and \eqref{eq:FMP}.
Note that this conclusion is quite general:
We did not explicitly 
specify $q$, $B_q$ and $M_1$
in the above arguments.
Their influence enters only through 
the observationally allowed range of $F$
for these parameters,
which generically satisfies \eqref{eq:FMP}.

\section{Summary and discussions}\label{sec:summary}

In this article, 
we constructed 
(de)constructed models of extra-natural inflation
which successfully explain the CMB observations.
We overcome the obstacle for (de)constructing
extra-natural inflation pointed out in 
\cite{ArkaniHamed:2003wu}
by introducing multiple (de)constructed extra dimensions,
building on our previous work
\cite{Furuuchi:2020klq}.
We compared the models with explicit choice of 
charged matter contents with CMB observations,
and derived the constraints on the model parameters.
The models were observationally viable
in a region of the parameter space.
We confirmed that the models 
successfully achieved
the trans-Planckian inflaton field excursion.
We also examined the mechanism of particle production
during inflation which may leave features
in primordial density perturbation
\cite{Furuuchi:2015foh,Furuuchi:2020klq}.
Under the natural and quite general assumptions,
we showed that the primordial features
from 
our (de)constructed extra-natural inflation models
would not be detectable in cosmological observations.

The natural inflation model 
with a single sinusoidal potential
is not favored
by the latest CMB observations \cite{Akrami:2018odb}.
However, simple modifications of 
the single sinusoidal potential may improve the fit 
to the observational data.
In this article, we studied
multi-natural inflation proposed in \cite{Czerny:2014wza}
as such an improved model.
Our (de)construction model provides a microscopic
theory of multi-natural inflation.
It will be interesting to explore
other modifications of the inflaton potential
from the simplest single sinusoidal potential
which can arise from (de)construction.

It will also be interesting 
to explore 
(de)constructed extra-natural inflation models
which predict detectable primordial features.
Examining the quite general assumptions
we have made to show that 
our models 
do not produce detectable primordial features
may provide a starting point for finding such models.

In this article, we restrict ourselves
to the regime where the number 
of the lattice points
in each direction is large.
In this regime, 
the resulting inflaton potential
coincides with that from
ordinary extra dimensions
in the leading order,
although 
the microscopic models 
do have different model parameters
with their own range of applicability.
In the meantime,
the difference between the ordinary extra dimensions
and the (de)constructed extra dimensions
becomes sharper when
the number of the lattice points in each direction
is small.
Thus it will be interesting to explore 
the regime in which 
the number of the lattice points 
in each direction is small.

\vskip8mm
\begin{center}
\textit{Acknowledgments}
\end{center}
\vskip-2mm
This work is supported in part by 
the Science and Engineering Research Board,
Department of Science and Technology, Government of India
under the project file number EMR/2015/002471.
The work of Suvedha Suresh Naik
is supported by
Dr.~T.M.A.~Pai PhD scholarship program of 
Manipal Academy of Higher Education.
Manipal Centre for
Natural Sciences, Centre of Excellence,
Manipal Academy of Higher Education 
is acknowledged for facilities and support.

\appendix
\section{Discrete Fourier Transform}\label{App:DFT}

We follow the same convention 
for the Discrete Fourier Transform (DFT)
as used in \cite{Furuuchi:2020klq}.
It is reviewed here for the convenience of the readers.

Let us first consider the DFT 
in one-dimensional periodic lattice.
Let us consider a cyclically ordered $N$ points
labeled by $j$ 
($j = 0, 1, \cdots , N-1$ $(\mathrm{mod}\,\, N)$).
Consider 
a variable ${\phi}_j$ which has a value on each point.
We use the following convention for
the discrete Fourier expansion of the variable
${\phi}_j$:
\begin{align}
{\phi}_{j} 
&= 
\frac{1}{\sqrt{N}}
\sum_{n=-\frac{N-1}{2}}^{\frac{N-1}{2}}
\tilde{{\phi}}_n\, e^{i\frac{2\pi  n j}{N}} 
\qquad (N: \mbox{odd})\,.
\label{eq:fnodd}
\\
{\phi}_{j} 
&= 
\frac{1}{\sqrt{N}}
\sum_{n=-\frac{N}{2}-1}^{\frac{N}{2}-1}
\tilde{{\phi}}_n\, e^{i\frac{2\pi  n j}{N}}
+
\frac{1}{\sqrt{N}}
\tilde{{\phi}}_{\frac{N}{2}}
(-)^j  
\qquad (N: \mbox{even})\,.
\label{eq:fneven}
\end{align}
Our convention is convenient
since when applied in (de)construction,
each KK mode is canonically normalized.

When ${\phi}_j$ is a real variable,
$\tilde{{\phi}}_{-n}^\ast = \tilde{f}_n$.
The
orthogonality 
of the exponential function:
\begin{equation}
\sum_{j=0}^{N-1}
\left(
e^{i\frac{2\pi  n_1 j}{N}} 
\right)^\ast
e^{i\frac{2\pi  n_2 j}{N}}
= N \delta_{n_1 n_2}\, ,
\label{eq:DFTo}
\end{equation}
leads to the following formula for the
discrete Fourier coefficient:
\begin{equation}
\tilde{{\phi}}_n
=
\frac{1}{\sqrt{N}}
\sum_{j=0}^{N-1}
{\phi}_j e^{-i\frac{2\pi  n j}{N}}\,.
\label{eq:invDFT}
\end{equation}

The generalization of the DFT
to $d$-dimensional periodic lattice is straightforward:
One just need to repeat the same procedure as above 
for each direction.
Let $N_I$ be the number of the points
in $I$-th direction ($I=1,2,\cdots,d$).
The discrete Fourier expansion is given as
\begin{equation}
{\phi}_{(\vec{j})} 
= 
\frac{1}{\prod_{I=1}^d {N_I}^{{1}/{2}}}
\sum_{n_1}
\sum_{n_2}
\cdots
\sum_{n_d}
\tilde{{\phi}}_{(\vec{n})}\, 
e^{i \sum_{J=1}^d \frac{2\pi  n_J j_J}{N_J}} 
\,.
\label{eq:invDFTd}
\end{equation}
Here,
$\vec{j}$ is a vector whose $I$-th component is $j_I$
($I=1,2,\cdots,d$),
and
$\vec{n}$ is a vector whose $I$-th component is $n_I$
($I=1,2,\cdots,d$).
Each sum over $n_I$ ($I=1,2,\cdots,d$)
follows the convention 
\eqref{eq:fnodd} or \eqref{eq:fneven}.

The Fourier coefficients are given as
\begin{equation}
\tilde{{\phi}}_{(\vec{n})}
=
\frac{1}{\prod_{I=1}^d {N_I}^{1/2}}
\sum_{j_1}
\sum_{j_2}
\cdots
\sum_{j_d}
{\phi}_{(\vec{j})} 
e^{-i \sum_{J=1}^d \frac{2\pi n_J j_J}{N_J}}\,.
\label{eq:FCDFTd}
\end{equation}
We call $\vec{n}=\vec{0}$ component
of the Fourier coefficients
``zero-mode.''
Explicitly,
\begin{equation}
\tilde{{\phi}}_{(\vec{0})}
=
\frac{1}{\prod_{I=1}^d {N_I}^{{1}/{2}}}
\sum_{j_1}
\sum_{j_2}
\cdots
\sum_{j_d}
{\phi}_{(\vec{j})} 
\,.
\label{eq:zeromode}
\end{equation}
When considering the discrete version of 
dimensional reduction,
it is useful to know the
value of ${\phi}_{(\vec{j})}$
when all the Fourier coefficients
except the zero-mode are zero:
\begin{equation}
{\phi}_{(\vec{j})}
\Bigl|_{\tilde{\phi}_{(\vec{n})} =0 
\,\, \mathrm{except}\,\, \vec{n}=\vec{0}}
=
\frac{1}{\prod_{{I}=1}^d N_{I}^{1/2}}
\tilde{\phi}_{(\vec{0})}
\,.
\label{eq:phi0}
\end{equation}

\section{One-loop effective 
potential}\label{app:Voneloop}

In this appendix, 
we derive the one-loop effective potential
for the zero-mode.
While in the case of $d=1$,
it is possible to write down 
the one-loop effective potential 
applicable for arbitrary number of
the lattice points
in a relatively simple form \cite{ArkaniHamed:2001nc},
we did not find such a simple expression for $d \geq 2$.
Therefore,
we will derive the one-loop effective potential 
in the leading order
in the number of the lattice points
following \cite{Furuuchi:2011px}.
The result formally coincides with the 
case of continuum extra dimensions
\cite{
Antoniadis:1990ew,Hatanaka:1998yp}.

Let us denote the 
contribution of a charged scalar
with charge $q$ and mass $m$
to the one-loop effective potential for 
the zero-modes $\tilde{A}_{(\vec{0})}^I$:
\begin{equation}
V^{q}_{\mathrm{1-loop}}(\tilde{A}_{(\vec{0})}^I)
=
\sum_{n_1}
\sum_{n_2}
\cdots
\sum_{n_d}
\int \frac{d^4k}{(2\pi)^4}
\ln 
\left[ 
k^2 + m^2 + \sum_{I=1}^d M_{n_I}^2(q, \tilde{A}_{(\vec{0})}^I)
\right]
\,,
\label{eq:Voneloop}
\end{equation}
where
\begin{equation}
M_{n_I}^2 (q,\tilde{A}_{(\vec{0})}^I)
:=
\frac{2}{a_I^2}
\left(
1 
- 
\cos 
\left[
\frac{q \tilde{A}_{(\vec{0})}^I}{{\cal F}_I}
+
\frac{2\pi n_I}{N_I}
\right]
\right)\,,
\label{eq:AppMq}
\end{equation}
and 
\begin{equation}
{\cal F}_I := f_I \prod_{J=1}^d N_J^{1/2} \,.
\label{eq:FI}
\end{equation}
In the above, we have 
analytically continued to 
the Euclidean time.
In \eqref{eq:AppMq}, for simplicity,
we restrict ourselves
to the universality restoration point
$\gamma_I = g^2$ $(I=1,\cdots,d)$,
at which the KK mass spectra of 
the charged scalar fields coincide
with that of the gauge field 
\cite{Furuuchi:2020klq}. 

It is convenient to define
\begin{equation}
\zeta_q(s)
:=
\sum_{n_1}
\sum_{n_2}
\cdots
\sum_{n_d}
\int \frac{d^4k}{(2\pi)^4}
\left[
k^2 + m^2 + \sum_{I=1}^d M_{n_I}^2(q, \tilde{A}_{(\vec{0})}^I)
\right]^{-s}
\,.
\label{eq:zeta}
\end{equation}
Using \eqref{eq:zeta}, 
the one-loop effective potential
\eqref{eq:Voneloop}
can be written as
\begin{equation}
V_{\mathrm{1-loop}}^q(\tilde{A}_{(\vec{0})}^I)
=
-
\frac{d \zeta_q(s)}{ds}
\Bigl.\Bigr|_{s=0}
\,.
\label{eq:Vzeta}
\end{equation}
We re-write \eqref{eq:zeta}
using the Schwinger parametrization:
\begin{align}
\zeta_q(s)
=
&
\sum_{n_1}
\sum_{n_2}
\cdots
\sum_{n_d}
\int \frac{d^4k}{(2\pi)^4}
\nn\\
&\frac{1}{\Gamma(s)}
\int_0^\infty
d\tau
\tau^{s-1}
\exp
\left[
-
\tau
\left(
k^2 + m^2 + \sum_{I=1}^d M_{n_I}^2(q, \tilde{A}_{(\vec{0})}^I)
\right)
\right]\,.
\label{eq:Schp}
\end{align}
After performing the Gaussian integral of $k$,
we obtain
\begin{equation}
\zeta_q(s)
=
\sum_{n_1}
\sum_{n_2}
\cdots
\sum_{n_d}
\frac{1}{2(4\pi)^2}
\frac{1}{\Gamma(s)}
\int_0^\infty
d\tau
\tau^{s-3}
\exp
\left[
-
\tau
\left(
m^2 + \sum_{I=1}^d M_{n_I}^2(q, \tilde{A}_{(\vec{0})}^I)
\right)
\right]\,.
\label{eq:Gaussdone}
\end{equation}
Then, its derivative 
with respect to $s$ gives
\begin{equation}
\frac{d\zeta_q(s)}{ds}
\Bigl.\Bigr|_{s=0}
=
\sum_{n_1}
\sum_{n_2}
\cdots
\sum_{n_d}
\frac{1}{2(4\pi)^2}
\int_0^\infty
d\tau
\tau^{s-3}
\exp
\left[
-
\tau
\left(
m^2 + \sum_{I=1}^d M_{n_I}^2(q, \tilde{A}_{(\vec{0})}^I)
\right)
\right]
\Biggl.\Biggr|_{s=0}
\,.
\label{eq:dzetads}
\end{equation}
In the above, 
we have used $\Gamma(s) \simeq 1/s\, +$ (finite) 
in the limit $s \rightarrow 0$,
and took the leading order term 
in the equation
anticipating the limit 
$s \rightarrow 0$.

Substituting \eqref{eq:dzetads}
into \eqref{eq:Vzeta} gives
\begin{align}
&V_{\mathrm{1-loop}}^q(\tilde{A}_{(\vec{0})}^I)\nn\\
&=
-
\sum_{n_1}
\sum_{n_2}
\cdots
\sum_{n_d}
\frac{1}{2(4\pi)^2}
\int_0^\infty
d\tau
\tau^{-3}
\exp
\left[
-
\tau
\left(
m^2 + \sum_{I=1}^d M_{n_I}^2(q, \tilde{A}_{(\vec{0})}^I)
\right)
\right]
\nn\\
&=
-
\sum_{n_1}
\sum_{n_2}
\cdots
\sum_{n_d}
\frac{1}{2(4\pi)^2}
\int_0^\infty
d\tau
\tau^{-3}
\exp
\left[
-
\tau
\left(
m^2 + \sum_{I=1}^d \frac{2}{a_I^2}
\right)
+ 
\tau
\sum_{I=1}^d \frac{2}{a_I^2}
\cos 
\left[
\frac{q \tilde{A}_{(\vec{0})}^I}{{\cal F}_I}
+
\frac{2\pi n_I}{N_I}
\right]
\right]
\,.
\nn\\
\label{eq:V1tau}
\end{align}
Using the following identity for 
the modified Bessel function $I_\nu(z)$
with integer $\nu$:
\begin{equation}
e^{z \cos \theta}
=
I_0(z)
+
2\sum_{\nu=1}
I_{\nu}(z) \cos \theta \,,
\label{eq:modB}
\end{equation}
we can rewrite \eqref{eq:V1tau} as
\begin{align}
&V_{\mathrm{1-loop}}^q(\tilde{A}_{(\vec{0})}^I)\nn\\
=&
-
\sum_{n_1}
\sum_{n_2}
\cdots
\sum_{n_d}
\frac{1}{2(4\pi)^2}
\int_0^\infty
d\tau
\tau^{-3}
\exp
\left[
-
\tau
\left(
m^2 + \sum_{I=1}^d \frac{2}{a_I^2}
\right)
+
\tau
\sum_{I=1}^d \frac{2}{a_I^2}
\cos
\left[
\frac{q \tilde{A}_{(\vec{0})}^I}{{\cal F}_I}
+
\frac{2\pi n_I}{N_I}
\right]
\right]
\nn\\
=&
-
\sum_{\ell_1=0}^{\infty}
\sum_{\ell_2=0}^\infty
\cdots
\sum_{\ell_d=0}^\infty
\frac{1}{2(4\pi)^2}
\int_0^\infty
d\tau
\tau^{-3}
\exp
\left[
-
\tau
\left(
m^2 + \sum_{I=1}^d \frac{2}{a_I^2}
\right)
\right]
\nn\\
&
\quad
\left\{
2^d
\prod_{I=1}^d
N_I
I_{N_I\ell_I} 
\left(
\frac{2\tau}{a_I^2}
\right)
\cos 
\left[
\frac{N_I \ell_I q \tilde{A}_{(\vec{0})}^I}{{\cal F}_I}
\right]
\right\}
\,.
\label{eq:V1nu}
\end{align}
On the other hand, 
we can expand the one-loop effective potential
in Fourier mode with respect to $\tilde{A}_{(\vec{0})}^I$
$(I = 1,2,\cdots, d)$:
\begin{equation}
V_{\mathrm{1-loop}}^q(\tilde{A}_{(\vec{0})}^I)
=
\sum_{\ell_1=0}^\infty
\sum_{\ell_2=0}^\infty
\cdots
\sum_{\ell_d=0}^\infty
V_{\vec{\ell}}
\prod_{I=1}^d
\cos 
\left[
\frac{q \ell_I N_I \tilde{A}_{(\vec{0})}^I}{{\cal F}_I}
\right]\,.
\label{eq:Vex}
\end{equation}
Here, 
$\vec{\ell}$ is a vector whose
$I$-th component is $\ell_I$.

Let us first analyze the simpler case $m^2=0$.
Using the 
integral representation of 
the modified Bessel function
with integer $\nu$
following from \eqref{eq:modB}:
\begin{equation}
I_\nu (z)
=
\frac{1}{\pi}
\int_0^\pi
d\theta \,
e^{z \cos \theta}
\cos (\nu \theta) \,,
\label{eq:intmodB}
\end{equation}
and taking 
$N_I$ large with
$N_I a_I = 2\pi L_I$ fixed,%
\footnote{
When $N_I \geq 3$, the limit $s \rightarrow 0$
does not lead to a divergence
other than the constant term, 
which we fine-tune
\cite{ArkaniHamed:2001nc}.
This result has already been used in \eqref{eq:V1tau}.}
we obtain
\begin{align}
 &V_{\vec{\ell}} \nn\\
=&
-
\frac{1}{2(4\pi)^2}
\left(
\frac{2}{\pi}
\right)^d
\int_0^\infty \frac{d\tau}{\tau^3}
\prod_{I=1}^d
\Biggl\{
N_I
\int_0^\pi d\theta_I
\exp
\left[
-\frac{2 N_I^2\tau}{(2 \pi L_I)^2}
\left(
1-\cos \theta_I
\right)
\right]
\cos 
\left[
N_I \ell_I \theta_I
\right]
\Biggr\}
\nn\\
=&
-
\frac{1}{2\cdot 2^d(4\pi)^2}
\left(
\frac{2}{\pi}
\right)^d
\int_0^\infty \frac{d\tau}{\tau^3}
\prod_{I=1}^d
\Biggl\{
N_I
\int_0^\pi d\theta_I
\nn\\
&\qquad \qquad
\Biggl(
\exp
\left[
-\frac{\tau}{(2\pi L_I)^2}
\left(
N_I^2 \theta_I^2
+
i \frac{(2\pi L_I)^2}{\tau} N_I \ell_I \theta_I
\right)
\right]
\nn\\
&
\qquad \qquad
+
\exp
\left[
-\frac{\tau}{(2\pi L_I)^2}
\left(
N_I^2 \theta_I^2
-
i \frac{(2\pi L_I)^2}{\tau} N_I \ell_I \theta_I
\right)
\right]
\Biggr)
\Biggr\}
+
\Ord (N_I^{-2})
\nn\\
=&
-
\frac{1}{2\cdot 2^d(4\pi)^2}
\left(
\frac{2}{\pi}
\right)^d
\int_0^\infty \frac{d\tau}{\tau^3}
\prod_{I=1}^d
\Biggl\{
\int_0^{N_I \pi} d \tilde{\theta}_I
\Biggl(
\exp
\left[
-\frac{\tau}{(2\pi L_I)^2}
\left(
\tilde{\theta}_I^2
+
i \frac{(2\pi L_I)^2}{\tau} \ell_I \tilde{\theta}_I
\right)
\right]
\nn\\
&
\qquad \qquad
+
\exp
\left[
-\frac{\tau}{(2\pi L_I)^2}
\left(
\tilde{\theta}_I^2
-
i \frac{(2\pi L_I)^2}{\tau} \ell_I \tilde{\theta}_I
\right)
\right]
\Biggr)
\Biggr\}
+
\Ord (N_I^{-2})
\quad (\tilde{\theta}_I = N_I \theta_I)
\nn\\
=&
-
\frac{1}{2 (4\pi)^2}
\left(
\frac{2}{\pi}
\right)^d
\int_0^\infty \frac{d\tau}{\tau^3}
\prod_{I=1}^d
\Biggl\{
\left(
\frac{\pi(2\pi L_I)^2}{\tau}
\right)^{1/2}
\exp
\left[
-\frac{(2\pi L_I)^2 \ell_I^2}{4\tau}
\right]
\Biggr\}
\nn\\
&\qquad \qquad
+
\Ord (N_I^{-2})
+
\Ord
\left(
N_I^{-1} e^{-(\pi N_I)^2}
\right)
\nn\\
=&
-
\frac{1}{2 (4\pi)^2}
\frac{2^d}{\pi^{d/2}}
\int_0^\infty d\tilde{\tau} \tilde{\tau}^{1+\frac{d}{2}}
\prod_{I=1}^d
\Biggl\{
(2\pi L_I)
\exp
\left[
-\frac{(2\pi L_I)^2 \ell_I^2 \tilde{\tau}}{4}
\right]
\Biggr\}
\nn\\
&\qquad \qquad
+
\Ord (N_I^{-2})
+
\Ord
\left(
N_I^{-1} e^{-(\pi N_I)^2}
\right)
\quad
\left(
\tilde{\tau} = \frac{1}{\tau}
\right)
\nn\\
=&
-
\frac{1}{2(4\pi)^2}
\frac{2^d}{\pi^{d/2}}
\prod_{I=1}^d
(2\pi L_I)
\left(
\frac{4}{\sum_{I=1}^d (2\pi L_I)^2 \ell_I^2}
\right)^{2+\frac{d}{2}}
\int_0^\infty dt \, t^{1+\frac{d}{2}}
e^{-t}
\nn\\
&\qquad \qquad
+
\Ord (N_I^{-2})
+
\Ord
\left(
N_I^{-1} e^{-(\pi N_I)^2}
\right)
\quad
\left(
\frac{\sum_{I=1}^d(2\pi L_I)^2 \ell_I^2 }{4}\, \tilde{\tau} = t
\right)
\nn\\
=&
-
\frac{1}{2(4\pi)^2}
\frac{2^d}{\pi^{d/2}}
\Gamma \left( 2 +\frac{d}{2} \right)
\prod_{I=1}^d
(2\pi L_I)
\left(
\frac{4}{\sum_{I=1}^d (2\pi L_I)^2 \ell_I^2}
\right)^{2+\frac{d}{2}}
+
\Ord (N_I^{-2})
+
\Ord
\left(
N_I^{-1} e^{-(\pi N_I)^2}
\right)\,.
\label{eq:Nlead}
\end{align}
In the above, we have used
\begin{equation}
\mathrm{erfc}\, x
:=
\frac{2}{\sqrt{\pi}}
\int_x^\infty dt\,
e^{-t^2}
=
\frac{e^{-x^2}}{x\sqrt{\pi}}
\sum_{n=0}^\infty
\frac{(-)^n (2n-1)!!}{(2x^2)^n}
\,.
\label{eq:erfc}
\end{equation}

Next we turn to the case $m^2\ne 0$.
\begin{align}
 &V_{\vec{\ell}}\nn\\
=&
-
\frac{1}{2(4\pi)^2}
\left(
\frac{2}{\pi}
\right)^d
\int_0^\infty \frac{d\tau}{\tau^3}
e^{-\tau m^2}
\prod_{I=1}^d
\Biggl\{
N_I
\int_0^\pi d\theta_I
\exp
\left[
-\frac{2 N_I^2\tau}{(2 \pi L_I)^2}
\left(
1-\cos \theta_I
\right)
\right]
\cos 
\left[
N_I \ell_I \theta_I
\right]
\Biggr\}
\nn\\
=&
-
\frac{1}{2\cdot 2^d (4\pi)^2}
\left(
\frac{2}{\pi}
\right)^d
\int_0^\infty \frac{d\tau}{\tau^3}
e^{-\tau m^2}
\prod_{I=1}^d
\Biggl\{
N_I
\int_0^\pi d\theta_I
\nn\\
&\qquad \qquad
\Biggl(
\exp
\left[
-\frac{\tau}{(2\pi L_I)^2}
\left(
N_I^2 \theta_I^2
+
i \frac{(2\pi L_I)^2}{\tau} N_I \ell_I \theta_I
\right)
\right]
\nn\\
&
\qquad \qquad
+
\exp
\left[
-\frac{\tau}{(2\pi L_I)^2}
\left(
N_I^2 \theta_I^2
-
i \frac{(2\pi L_I)^2}{\tau} N_I \ell_I \theta_I
\right)
\right]
\Biggr)
\Biggr\}
+
\Ord (N_I^{-2})
\nn\\
=&
-
\frac{1}{2 \cdot 2^d (4\pi)^2}
\left(
\frac{2}{\pi}
\right)^d
\int_0^\infty \frac{d\tau}{\tau^3}
e^{-\tau m^2}
\prod_{I=1}^d
\Biggl\{
\int_0^{N_I \pi} d \tilde{\theta}_I
\Biggl(
\exp
\left[
-\frac{\tau}{(2\pi L_I)^2}
\left(
\tilde{\theta}_I^2
+
i \frac{(2\pi L_I)^2}{\tau} \ell_I \tilde{\theta}_I
\right)
\right]
\nn\\
&
\qquad \qquad
+
\exp
\left[
-\frac{\tau}{(2\pi L_I)^2}
\left(
\tilde{\theta}_I^2
-
i \frac{(2\pi L_I)^2}{\tau} \ell_I \tilde{\theta}_I
\right)
\right]
\Biggr)
\Biggr\}
+
\Ord (N_I^{-2})
\quad
(\tilde{\theta}_I = N_I \theta_I)
\nn\\
=&
-
\frac{1}{2(4\pi)^2}
\left(
\frac{2}{\pi}
\right)^d
\int_0^\infty \frac{d\tau}{\tau^3}
e^{-\tau m^2}
\prod_{I=1}^d
\Biggl\{
\left(
\frac{\pi(2\pi L_I)^2}{\tau}
\right)^{1/2}
\exp
\left[
-\frac{(2\pi L_I)^2 \ell_I^2}{4\tau}
\right]
\Biggr\}
\nn\\
&\qquad \qquad
+
\Ord (N_I^{-2})
+
\Ord
\left(
N_I^{-1} e^{-(\pi N_I)^2}
\right)
\nn\\
=&
-
\frac{1}{2(4\pi)^2}
\frac{2^d}{\pi^{d/2}}
\int_0^\infty d\tilde{\tau} 
e^{-\frac{m^2}{\tilde{\tau}}}
\tilde{\tau}^{1+\frac{d}{2}}
\prod_{I=1}^d
\Biggl\{
(2\pi L_I)
\exp
\left[
-\frac{(2\pi L_I)^2 \ell_I^2 \tilde{\tau}}{4}
\right]
\Biggr\}
\nn\\
&\qquad \qquad
+
\Ord (N_I^{-2})
+
\Ord
\left(
N_I^{-1} e^{-(\pi N_I)^2}
\right)
\quad
\left(
\tilde{\tau} = \frac{1}{\tau}
\right)
\nn\\
=&
-
\frac{1}{2(4\pi)^2}
\frac{2^d}{\pi^{d/2}}
\prod_{I=1}^d
(2\pi L_I)
\left(
\frac{4}{\sum_{I=1}^d (2\pi L_I)^2 \ell_I^2}
\right)^{2+\frac{d}{2}}
\int_0^\infty dt \, t^{1+\frac{d}{2}}
e^{-t - \frac{z^2}{4t}}
\nn\\
&\qquad
+
\Ord (N_I^{-2})
+
\Ord
\left(
N_I^{-1} e^{-(\pi N_I)^2}
\right)
\quad
\left(
\frac{\sum_{I=1}^d(2\pi L_I)^2 \ell_I^2 }{4}\, \tilde{\tau} = t\,,
\,\,
z^2 = {m^2}{\sum_{I=1}^d(2\pi L_I)^2 \ell_I^2 }
\right)
\nn\\
=&
-
\frac{1}{2(4\pi)^2}
\frac{2^d}{\pi^{d/2}}
2 
\left(
\frac{z}{2}
\right)^{-2-\frac{d}{2}}
K_{-\left(2 +\frac{d}{2}\right)}(z)
\prod_{I=1}^d
(2\pi L_I)
\left(
\frac{4}{\sum_{J=1}^d (2\pi L_J)^2 \ell_J^2}
\right)^{2+\frac{d}{2}}
\nn\\
&\qquad \qquad
+
\Ord (N_I^{-2})
+
\Ord
\left(
N_I^{-1} e^{-(\pi N_I)^2}
\right)\,.
\label{eq:Nleadm}
\end{align}
In the above,
we have used 
the
integral representation of 
the modified Bessel function
\begin{equation}
\int_0^\infty
dt\,
e^{-t-\frac{z^2}{4t}}
t^{\nu-1}
=
2 
\left(
\frac{z}{2}
\right)^{\nu}
K_{-\nu} (z) \,,
\label{eq:Knu}
\end{equation}
which is valid for 
$|\arg z | < \frac{\pi}{4}$.

Using the limit
$z \rightarrow 0$,
\begin{equation}
\int_0^\infty
dt\,
e^{-t-\frac{z^2}{4t}}
t^{\nu-1}
\rightarrow
\Gamma(\nu) 
+
\Ord (z^2)
\,,
\label{eq:zzerolim}
\end{equation}
we recover the previous result 
\eqref{eq:Nlead}
for the case $m^2 =0$.
In the meantime, from the
asymptotic expansion 
of $K_\nu(z)$ for large $z$:
\begin{equation}
K_\nu (z)
\sim
\sqrt{\frac{\pi}{2z}}
e^{-z}
\sum_{n=0}^\infty
\frac{(\nu,n)}{(2z)^n}\,,
\label{eq:Knuasympt}
\end{equation}
where
\begin{align}
(\nu,n) 
&= \frac{\Gamma\left(\nu+n+\frac{1}{2}\right)}{n! \Gamma\left(\nu-n+\frac{1}{2}\right)}
\qquad (n\ne0)\,,
\nn\\
(\nu,0) &= 1\,,
\label{eq:nun}
\end{align}
we observe that when $m^2 \gg 1/L_I^2$,
the contribution to the one-loop effective potential
is exponentially suppressed.
Therefore, 
when 
calculating the one-loop effective potential,
we can safely neglect the contributions
from the fields which have mass above the KK-scale.
In the meantime, the contribution
of the fields which are
much lighter than the KK-scale
can be approximated
by the massless limit
using \eqref{eq:zzerolim}.

In this article, we set
\begin{align}
f_I &= f \qquad (\mbox{for all $I$})\,,
\label{eq:Appallf}
\\
N_I &= N \qquad (\mbox{for all $I \ne 1$})\,,
\label{eq:AppNIN}
\\
N &\gg N_1
\label{eq:AppNN1}
\,.
\end{align}
From
\eqref{eq:Appallf}, \eqref{eq:AppNIN} and \eqref{eq:LI},
all $L_I$ except $I=1$ are the same.
We denote $L_I = L$ for all $I \ne 1$.
As described in the main body, 
with the simplifying assumptions
\eqref{eq:Appallf} and \eqref{eq:AppNIN},
the condition \eqref{eq:AppNN1}
can be used to make
the potential 
such that
in the $\phi := \tilde{A}_{(\vec{0})}^1$ 
direction
satisfies the slow-roll condition
while the $\tilde{A}_{(\vec{0})}^I$ 
directions ($I \ne 1$)
do not.
Then, during inflation
we can safely set $\tilde{A}_{(\vec{0})}^I$ ($I \ne 1$) to its value
at the bottom of the potential: 
$\tilde{A}_{(\vec{0})}^I = 0$ for $I \ne 1$,
and the model is described as a single-field inflation model.

We approximate the one-loop effective potential of $\phi$ 
by taking only
$\vec{\ell} = (1,0,\cdots,0)$ term
in \eqref{eq:Vex},
since the remaining terms
rapidly decrease with $\ell_I$.\footnote{
The terms omitted here might have relevance in future observations,
for example, in the observations 
of the running of the spectral index or the further running of it
\cite{Kohri:2014rja}.}
We obtain
\begin{equation}
V_{\mathrm{1-loop}}^q (\tilde{A}_{(\vec{0})}^1 = \phi, \tilde{A}_{(\vec{0})}^I =0\, (I\ne 1))
\simeq
V^q (\phi)
:=
-
\Lambda^4 \cos \left[ \frac{q \phi}{F} \right]\,,
\label{eq:AppVq}
\end{equation}
where
\begin{align}
\Lambda^4
&=
\frac{1}{2(4\pi)^2}
\frac{2^d}{\pi^{d/2}}
\Gamma \left( 2 + \frac{d}{2} \right)
(2\pi L_1)(2\pi L)^{d-1}
\left(
\frac{4}{(2\pi L_1)^2}
\right)^{2+\frac{d}{2}}
\nn\\
&=
\frac{1}{2(4\pi)^2}
\frac{2^d}{\pi^{{d}/{2}}}
\Gamma\left( 2 + \frac{d}{2} \right)
N_1 N^{d-1} \left( \frac{2}{N_1} \right)^{4+d}
(gf)^4 
\,,
\label{eq:appLambda4}
\end{align}
and
\begin{equation}
F 
=
\frac{N^{\frac{d-1}{2}}f}{N_1^{\frac{1}{2}}}\,.
\label{eq:appFf}
\end{equation}

\section{Fitting Functions}\label{App:Fit}

As explained in the main body,
for generic values of parameters,
we cannot analytically perform integration 
in \eqref{eq:efolds}
to have explicit functional form
of ${\cal N}_\ast$ as a function of $\phi_\ast$,
or $\phi_\ast$ as a 
function of ${\cal N}_\ast$.
As a result, 
we do not have 
an explicit functional form $\Phi [F]$
given in \eqref{eq:Phi} or \eqref{eq:Phi2}.
This makes it hard to understand 
the dependence of the model predictions
on the parameter $F$ 
without relying on numerical tools.
To ease this issue, 
it is convenient to
have a fitting function $\Phi_{fit}[F]$
which approximates $\Phi [F]$
for the range of $F$ of interest.
For this purpose,
we first numerically evaluate the 
values 
of $\Phi [F]$ 
for the range of $F$ of interest
with step $0.1$. 
Then we fit the 
logarithm of these values
with a polynomial with degree two.
The exponential of this fitting polynomial 
can be used as an approximation to $\Phi [F]$. 

Similarly,
when $q$, $B_q$ 
and ${\cal N}_\ast$
are fixed,
$H_\ast := H(\phi_\ast)$ 
in \eqref{eq:COBE}
depends only on the parameter $F$.
We numerically evaluate values of
$H^2_\ast$
for the range of $F$ of interest
with step $0.1$, 
and fit these values
with a polynomial of degree two.

\subsubsection*{The model
\texorpdfstring{$q=2$}{q}, 
\texorpdfstring{$B_q=0.2$}{Bq}, 
\texorpdfstring{$M_1=5$}{M1} 
with 
\texorpdfstring{${\cal N}_\ast =60$}{Nast}}

\begin{figure}[H]
    \centering
    \includegraphics[width=9cm]{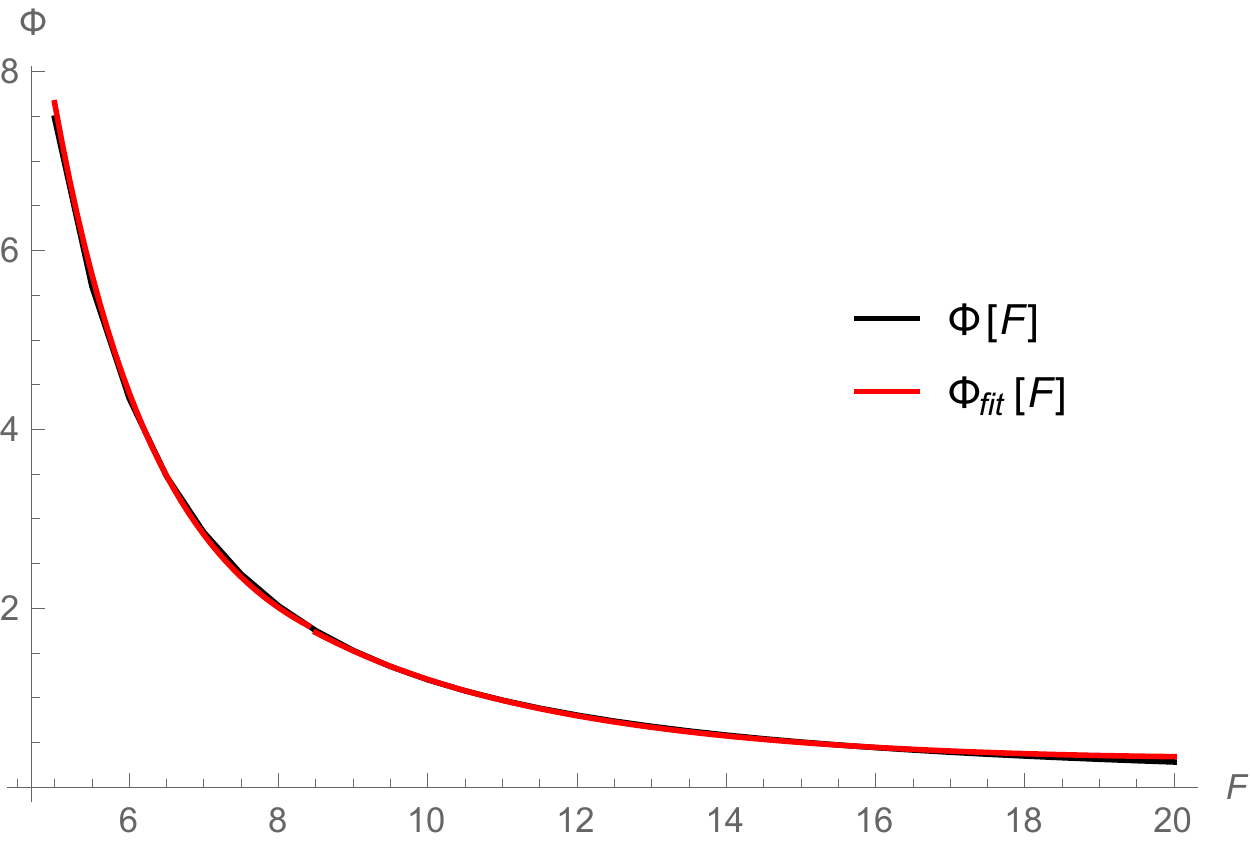}
    \caption{$\Phi [F]$ in \eqref{eq:Phi} 
		which is numerically 
    evaluated (black) and the fitting polynomial 
		$\Phi_{fit} [F]$ (red) are plotted 
		for a range of values of $F$
		for the model 
    $q=2$, $B_q=0.2$ 
		with ${\cal N}_\ast =60$.}
    \label{fig:AppPoly}
\end{figure}

\begin{figure}[H]
    \centering
    \includegraphics[width=9cm]{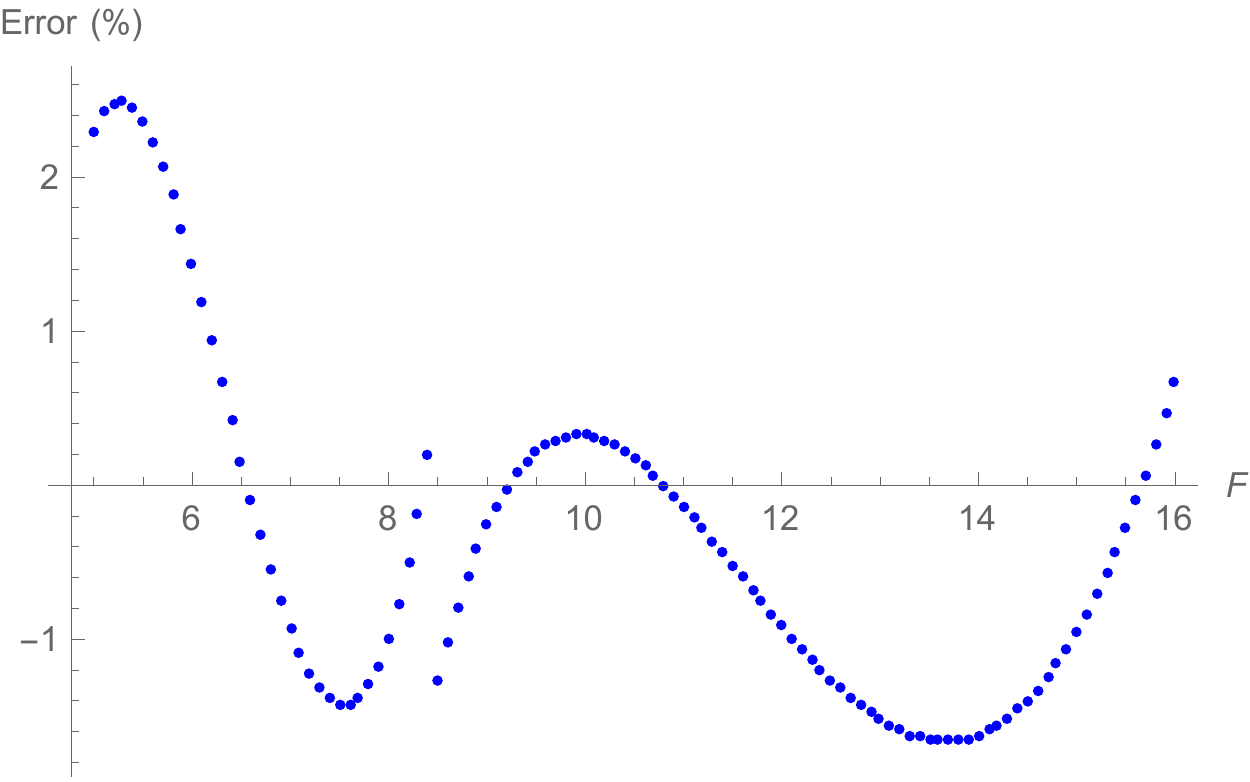}
    \caption{The plot of the error of the fitting 
		$(\Phi_{fit} [F]-\Phi [F]) / \Phi[F]$
		in percent ($\%$)
		for the model
    $q=2$, $B_q=0.2$ 
		with ${\cal N}_\ast =60$.}
    \label{fig:ApperrorPoly}
\end{figure}

The 
fitting function for $\Phi [F]$ 
we provide is 
\begin{equation}
    \Phi_{fit} [F] 
		=
		\begin{cases}
		\exp \left[ 6.37 - 1.13 F + 0.0526 F^2 \right] &(6.0 \leq F < 8.5) \, ,
		\\
		\exp \left[ 3.44 - 0.425 F + 0.00995 F^2 \right] & (8.5 \leq F \leq 16 ) \,.
		\end{cases}
\label{eq:Phifit}
\end{equation}
$\Phi [F]$ and $\Phi_{fit} [F]$
are plotted for the range of $F$ of interest 
in Fig.~\ref{fig:AppPoly}.
The error of the fit is 
plotted in Fig.~\ref{fig:ApperrorPoly}.
We observe that the size of the error is around $1\%$ or less
in the range of $F$ of interest.

\begin{figure}[H]
    \centering
    \includegraphics[width=9cm]{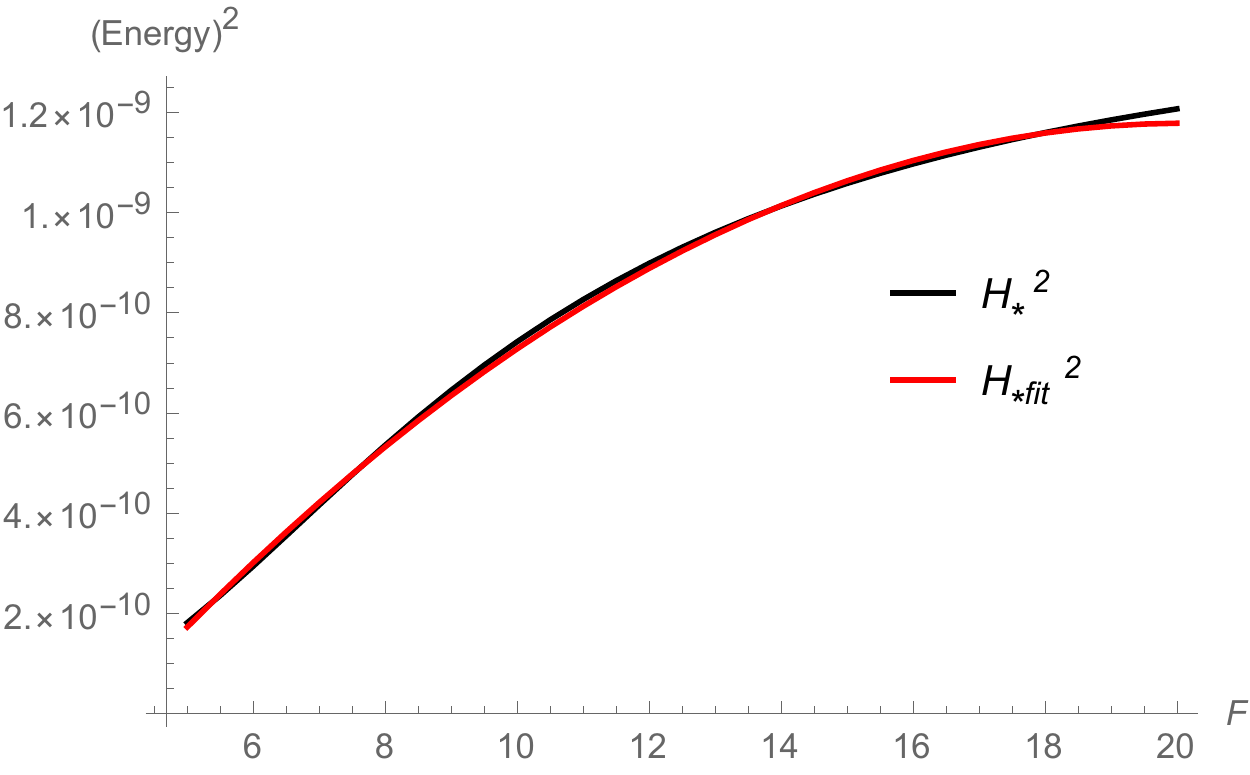}
    \caption{The plot of the square of the 
		Hubble parameter at the pivot scale $H_\ast^2$
		numerically evaluated (black) 
		for the model
    $q=2$, $B_q=0.2$, $M_1=5$ with ${\cal N}_\ast =60$
		and the fitting polynomial $H_{\ast fit}^2[F]$
    \eqref{eq:H2fit} (red)
		for the range of $F$ of interest
		$F_{l.b.} = 6.4 \leq F \leq F_{u.b.} = 16$.}
    \label{fig:AppHubbleparameter}
\end{figure}

\begin{figure}[H]
    \centering
    \includegraphics[width=9cm]{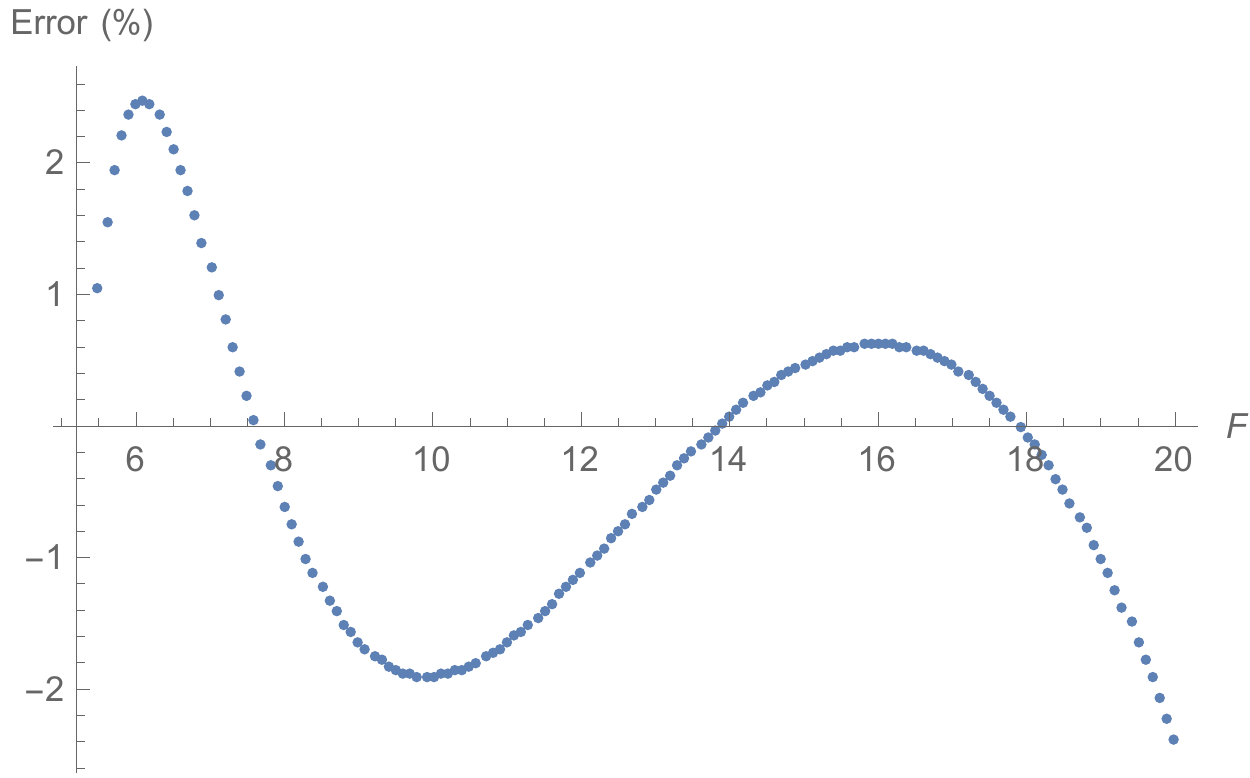}
    \caption{The plot of the error $(H_{\ast fit}^2 - H_\ast^2) / H_\ast^2$ 
		in percent ($\%$) for the model
    $q=2$, $B_q=0.2$, $M_1=5$ with ${\cal N}_\ast =60$
		for the range of $F$ of interest
		$F_{l.b.} = 6.4 \leq F \leq F_{u.b.} = 16$.}
    \label{fig:Apperrorhubble}
\end{figure}

The fitting function we provide for $H_\ast^2$ is
\begin{equation}
    H_{\ast fit}^{2} [F] 
		= 
		- 6.02\times 10^{-10} + 1.77 \times 10^{-10}F - 0.0116 F^2
		\,.
\label{eq:H2fit}
\end{equation}
$H_\ast^2$ and $H^2_{fit}$ are plotted in
Fig.~\ref{fig:AppHubbleparameter}.
The error 
of the fit
is plotted in Fig.~\ref{fig:Apperrorhubble}.
We observe that the size of the error is around $1\%$ or less
in the range of $F$ of interest.

\subsubsection*{The model
\texorpdfstring{$q=3$}{q}, 
\texorpdfstring{$B_q=0.25$}{Bq}, 
\texorpdfstring{$M_1=4$}{M1}
with 
\texorpdfstring{${\cal N}_\ast =60$}{Nast}}

\begin{figure}[H]
    \centering
    \includegraphics[width=9cm]{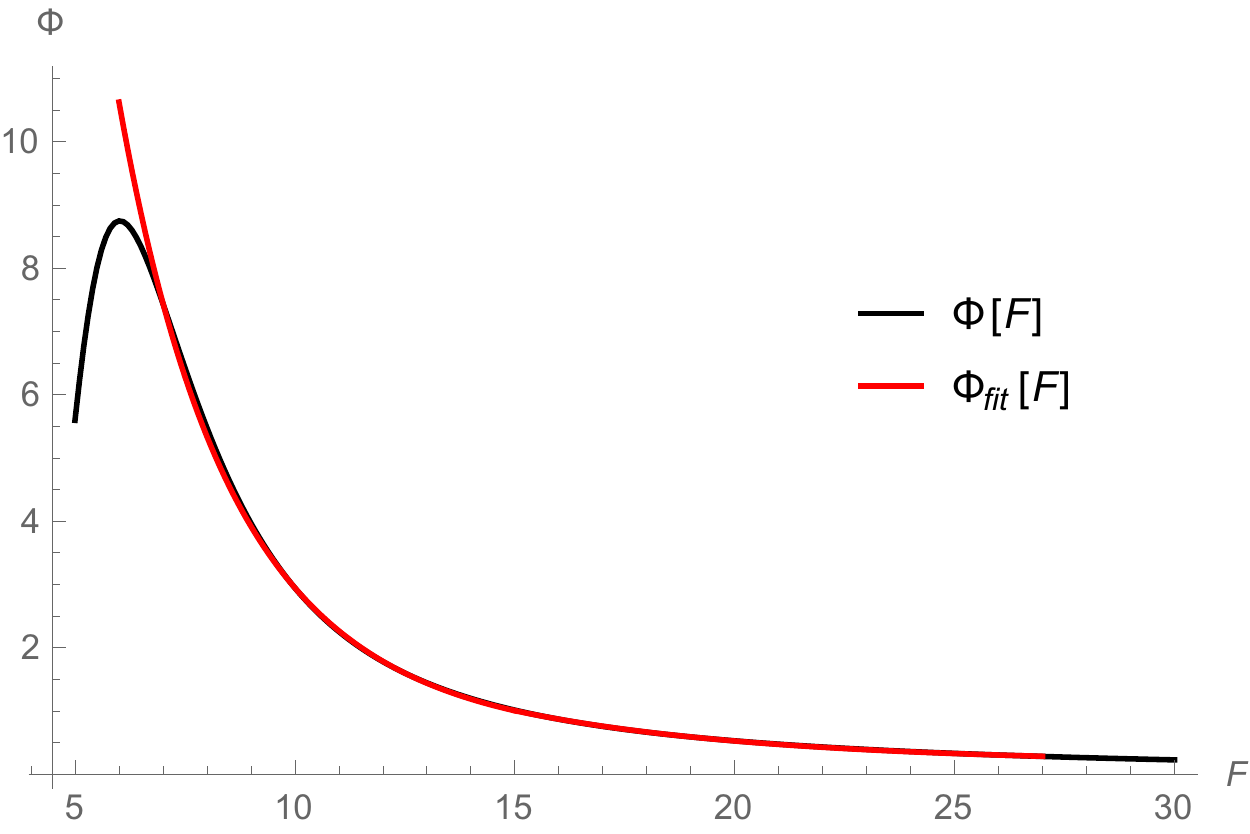}
    \caption{$\Phi [F]$ in \eqref{eq:Phi2} 
		which is numerically 
    evaluated (black) and the fitting polynomial 
		$\Phi_{fit} [F]$ (red) are plotted 
		for a range of values of $F$
		for the model 
    $q=3$, $B_q=0.25$ 
		with ${\cal N}_\ast =60$.}
    \label{fig:AppPoly2}
\end{figure}

\begin{figure}[H]
    \centering
    \includegraphics[width=9cm]{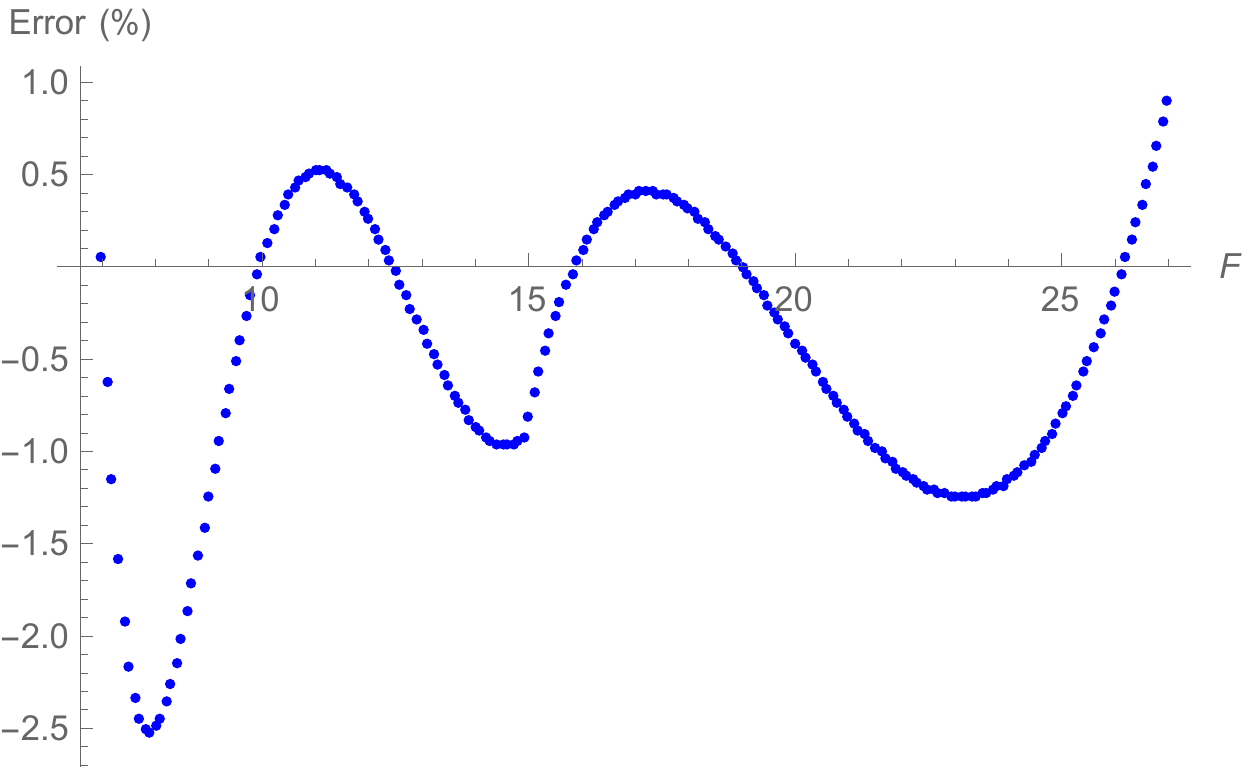}
    \caption{The plot of the error 
	  $(\Phi_{fit} [F]-\Phi [F]) /\Phi[F]$ in percent ($\%$)
		for the model
    $q=3$, $B_q=0.25$ 
		with ${\cal N}_\ast =60$.}
    \label{fig:ApperrorPoly2}
\end{figure}

The 
fitting function for $\Phi [F]$ 
we provide is 
\begin{equation}
\Phi_{fit}[F]
= 
\begin{cases}
\exp \left[ 4.99 - 0.508 F + 0.0117 F^2 \right] & (7 \leq F <  15)\,, \\
\exp \left[ 2.99 - 0.251 F + 0.00346 F^2 \right] & (15 \leq F \leq  27)\,.
\end{cases}
\label{eq:PhiFit2}
\end{equation}
$\Phi [F]$ and $\Phi_{fit} [F]$
are plotted for the range of $F$ of interest
in Fig.~\ref{fig:AppPoly2}.
The error of the fit is 
plotted in Fig.~\ref{fig:ApperrorPoly2}.
We observe that the size of the error is around $2\%$ or less
in the range of $F$ of interest.

\begin{figure}[H]
    \centering
    \includegraphics[width=9cm]{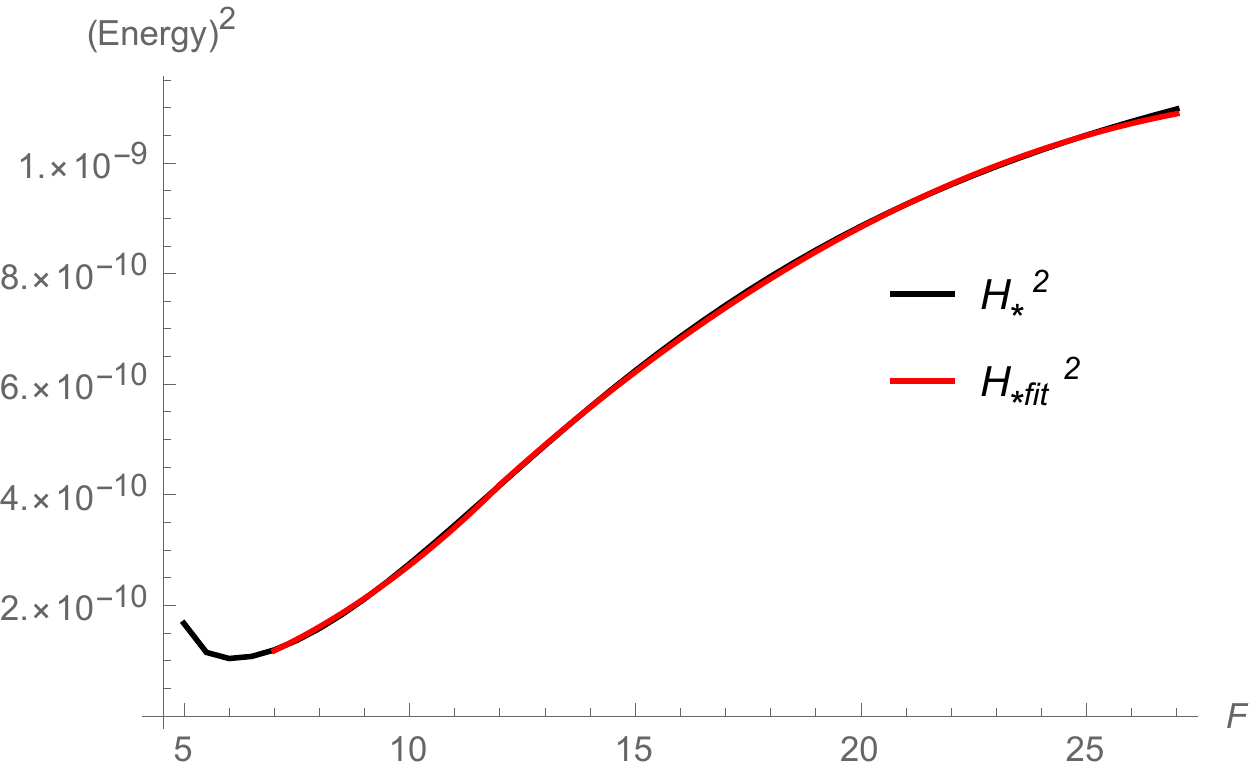}
    \caption{The plot of the square of the 
		Hubble parameter at the pivot scale $H_\ast^2$
		numerically evaluated 
		(black)
		for the model 
		$q=3$, $B_q=0.25$, $M_1=4$ with ${\cal N}_\ast =60$
		and the fitting polynomial $H^2_{\ast fit}$ 
		\eqref{eq:H2fit2} (red)
	  for the range of $F$ of interest
		$F_{l.b.} = 6.9 \leq F \leq F_{u.b.} = 26$.}
    \label{fig:AppHubbleparameter2}
\end{figure}

\begin{figure}[H]
    \centering
    \includegraphics[width=9cm]{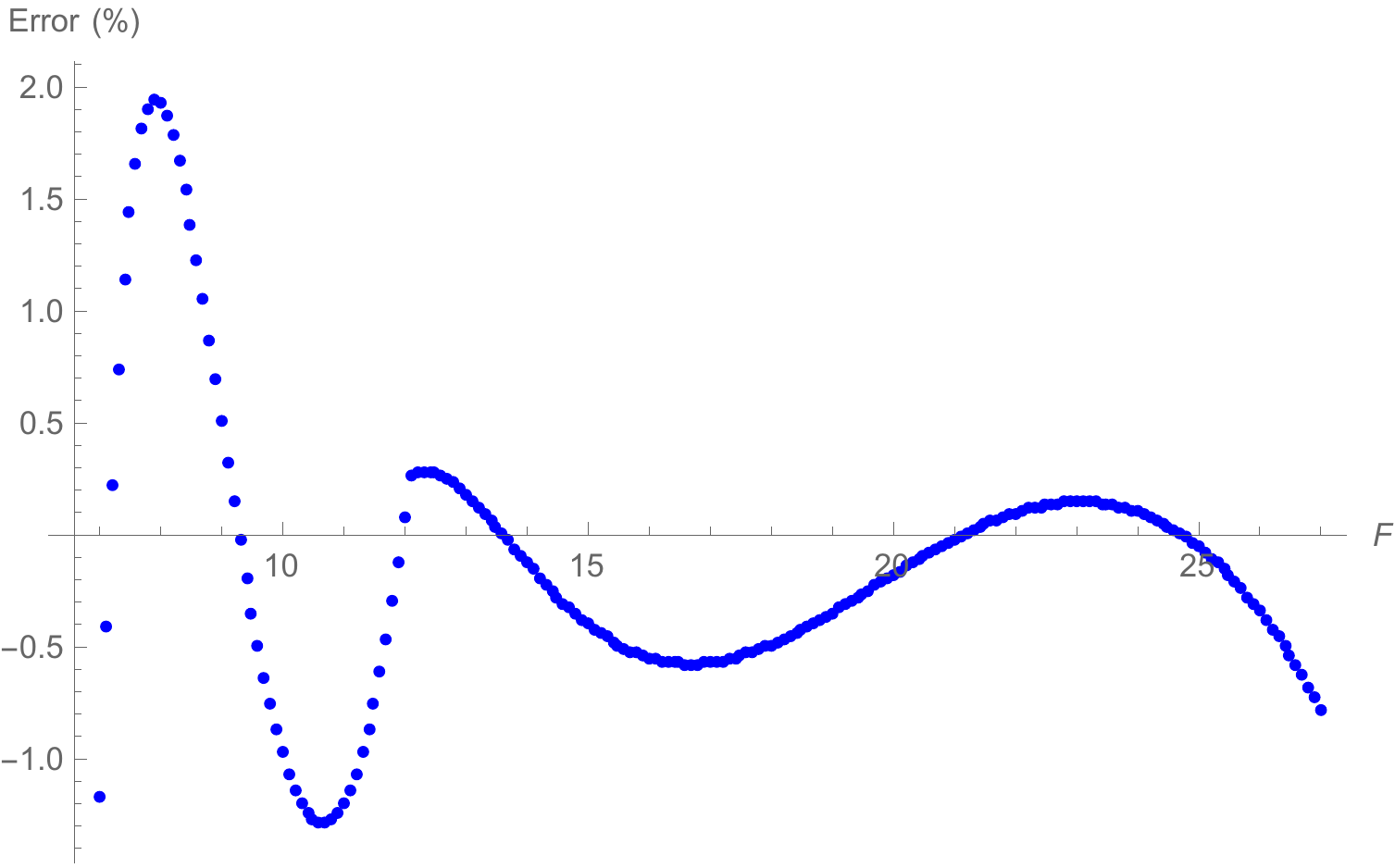}
    \caption{The plot of the error 
		$(H_{\ast fit}^2 - H_\ast^2) / H_\ast^2$ in percent ($\%$)
		for the model
    $q=3$, $B_q=0.25$, $M_1=4$ with ${\cal N}_\ast =60$
		for the range of $F$ of interest
		$F_{l.b.} = 6.9 \leq F \leq F_{u.b.} = 26$.}
    \label{fig:Apperrorhubble2}
\end{figure}

The fitting function we provide for $H_\ast^2$ is
\begin{equation}
    H_{\ast fit}^{2} [F] 
		= 
    \begin{cases}
		5.04\time 10^{-11}-1.97\times10^{-11} F + 4.19\times10^{-12} F^2
		& (7 \leq F \leq 12) \,,
		\\
		-7.44\time 10^{-10}-1.2\times10^{-10} F -1.93\times10^{-12} F^2
		& (12 < F \leq 27) \,.
		\end{cases}
\label{eq:H2fit2}
\end{equation}
$H_\ast^2$ and $H^2_{\ast fit}$ are plotted in
Fig.~\ref{fig:AppHubbleparameter2}.
The error of the fit
is plotted in Fig.~\ref{fig:Apperrorhubble2}.
We observe that the size of the error is around $2\%$ or less
in the range of $F$ of interest.

\clearpage 
\bibliography{ExNDec_Ref}
\bibliographystyle{utphys}
\end{document}